\documentclass{PoS}

\usepackage{varioref,exscale,latexsym,amsmath,amssymb}
\usepackage{graphicx}

\usepackage{bm}

\allowdisplaybreaks[1]

\def\sq[#1,#2]{\left[#1\,#2\right]}
\def\an[#1,#2]{\left\langle#1\,#2\right\rangle}
\def\spab[#1,#2,#3]{\left\langle#1|#2|#3\right]}

\def\beq{\begin{equation}}

\def\eeq{\end{equation}}

\def\beqa{\begin{eqnarray}}

\def\eeqa{\end{eqnarray}}

\title{Illuminating Light Bending}

\ShortTitle{Illuminating Light Bending}

\author{N. Emil J. Bjerrum-Bohr\\
         Niels Bohr International Academy and Discovery Center\\ The Niels Bohr Institute,
 Blegamsvej 17,
DK-2100 \\ Copenhagen \O, Denmark
}
\author{Barry R. Holstein\\
         Department of Physics-LGRT\\
University of Massachusetts\\
Amherst, MA  01003 US
}

\author{John F. Donoghue\\
         Department of Physics-LGRT\\
University of Massachusetts\\
Amherst, MA  01003 US
}

\author{Ludovic Plant\'e\\
    Institut de Physique Th\'eorique\\
CEA, IPhT, F-91191 Gif-sur-Yvette, France
}

\author{\speaker{Pierre Vanhove}%
       \thanks{IPHT-t15/011/, IHES/P/15/05.}\\
    Institut de Physique Th\'eorique\\
CEA, IPhT, F-91191 Gif-sur-Yvette, France
}

\abstract{
The interactions of gravitons with spin-1 matter are calculated in
parallel with the well known photon case.  It is shown that graviton
scattering amplitudes can be factorized into a product of familiar
electromagnetic forms, and cross sections for various reactions
are straightforwardly evaluated using helicity methods.  Universality relations are
identified.  Extrapolation to zero mass yields scattering amplitudes for
photon-graviton and graviton-graviton scattering. The phenomenon of light bending near a massive object, which is generally treated using classical general relativity, is discussed from alternative points of view.
}

\FullConference{Corfu Summer Institute 2016 "School and Workshops on Elementary Particle Physics and Gravity"\\
		 31 August - 23 September, 2016\\
		 Corfu, Greece}

\begin{document}
\section{Introduction}
The calculation of photon interactions with matter is part of
any introductory (or advanced) course on quantum mechanics.  Indeed
the evaluation of the Compton scattering cross section is a
standard exercise in relativistic quantum mechanics, since gauge
invariance together with the masslessness of the photon allow the
results to be presented in terms of relatively simple analytic
forms~\cite{hol}.

One might expect a similar analysis to be applicable to the
interactions of gravitons since, like photons, gravitons
are massless and subject to a gauge invariance.
Also, just as virtual photon exchange leads to a detailed
understanding of electromagnetic interactions between charged
systems, a careful treatment of virtual graviton exchange allows
an understanding not just of Newtonian gravity, but also of
spin-dependent phenomena---geodetic precession and Lense-Thirring
frame dragging---associated with general relativity which
have recently been verified by gravity probe
B~\cite{gpb}.  However, despite these parallels, examination
of quantum mechanics texts reveals that (with one
exception~\cite{ari}) the case of graviton interactions is {\it not}
discussed in any detail. There are at least three reasons for this
situation:
\begin{itemize}
\item [i)] the graviton is a spin-two particle, as opposed to the
spin-one photon, so that the interaction forms are more
complex, involving symmetric and traceless second rank tensors
rather than simple Lorentz four-vectors;
\item [ii)] there exist fewer experimental results with which to confront
the theoretical calculations. Fundamental questions
beyond the detection of quanta of gravitational fields  have been
exposed in~\cite{DYSON:2013jra};
\item [iii)] in order to guarantee gauge invariance one must include,
in many processes, the contribution
from a graviton pole term, involving a {\it triple-graviton coupling}.
This vertex is a sixth rank tensor and contains a multitude of
kinematic forms.
\end{itemize}

A century after the
classical theory of general relativity and Einstein's\footnote{{\sl
    Gleichwohl m\"u\ss ten die Atome zufolge
  der inneratomischen Elektronenbewegung nicht nur
  elektromagnetische, sondern auch Gravitationsenergie ausstrahlen,
  wenn auch in winzigem Betrage. Da dies in Wahrheit in der Natur
  nicht zutreffen d\"urfte, so scheint es, da\ss{} die Quantentheorie nicht
  nur die Maxwellsche Elektrodynamik, sondern auch die neue
  Gravitationstheorie wird modifizieren
  m\"ussen}} argument for a
quantization of gravity~\cite{Einstein:1916cc}, we are still seeking an experimental signature 
of quantum gravity effects. This
paper presents and extends recent works, where elementary quantum gravity
processes display new and very distinctive behaviors.

Recently, however, using powerful (string-based) techniques, which
simplify conventional quantum field theory calculations, it has
been demonstrated that the scattering of
gravitons from an elementary target of arbitrary spin
factorizes~\cite{fac}, a feature that had been noted ten years
previously by Choi et al. based on gauge theory
arguments~\cite{kor}. This factorization property, which is sometime
concisely described by the phrase ``gravity is the square of a gauge
theory", permits a relatively elementary evaluation of various
graviton amplitudes and opens the possibility of studying gravitational
processes in physics coursework.  In an earlier paper by one of us~\cite{ppa}
it was shown explicitly how, for both spin-0 and spin-${1\over 2}$ targets, the use of factorization enables
elementary calculation of both the graviton photoproduction,
$$\gamma+S\rightarrow g+S,$$
and gravitational Compton scattering,
$$g+S\rightarrow g+S,$$
reactions in terms of elementary photon reactions. This simplification means that graviton interactions
can now be discussed in a basic quantum
mechanics course and opens the possibility of treating interesting cosmological applications.

In the present paper we extend the work begun in~\cite{ppa} to the case of a spin-1
target and demonstrate and explain the origin of various universalities, {\it i.e.,} results which
are independent of target spin.  In
addition, by taking the limit of vanishing target mass we show how both graviton-photon
and graviton-graviton scattering may be determined using elementary methods.

In section 2 then, we review the electromagnetic interactions
of a spin one system.  In section 3 we calculate the ordinary Compton scattering cross section for a spin-1
target and compare with the analogous spin-0 and spin-${1\over 2}$ forms.  In section 4 we
examine graviton photoproduction and gravitational Compton scattering for a spin-1 target and again compare
with the analogous spin-0 and spin-${1\over 2}$ results. In section 5
we study the massless limit and show how both photon-graviton and graviton-graviton scattering
can be evaluated, resolving a subtlety which arises in the
derivation. Section~6 discusses some intriguing properties of the
forward cross-section.  In section~7 we review the classical physics calculation of the
bending of light, including both lowest order and next to
leading order corrections. After a derivation of the
gravitational interaction of massless and massive systems in section
8, in section 9 we present at an alternative derivation in terms of
geometrical optics, which uses the wave interpretation of light
propagation.  Then in section 10, we examine an additional way to
 derive the light bending, in terms of a quantum mechanical small
 angle scattering (eikonal) picture following the approach in~\cite{Bjerrum-Bohr:2016hpa}.  A brief concluding section summarizes our results.  The equivalence between results derived via these on the surface disparate techniques serves as an interesting example which can introduce students to new ways to analyze a familiar problem.
 Two appendices contain formalism and calculational details.

\section{Spin One Interactions: a Lightning Review}

We begin by reviewing the photon and graviton interactions of a spin-1 system. Recall that for a massive spin-0 system, we  
generate the photon interactions by writing down the free Lagrangian for a scalar field $\phi$
\begin{equation}
{\cal L}_0^{S=0}=\partial_\mu\phi^\dagger\partial^\mu\phi-m^2\phi^\dagger\phi\,,
\end{equation}
and making the minimal substitution~\cite{jdj}
$$i\partial_\mu\longrightarrow iD_\mu\equiv i\partial_\mu-eA_\mu\,.$$
This procedure leads to the familiar interaction Lagrangian
\begin{equation}
{\cal L}_{int}^{S=0}=-iA_\mu\phi^\dagger \overleftrightarrow{\partial}^\mu\phi+e^2A^\mu A^\nu\eta_{\mu\nu}\phi^\dagger\phi\,,
\end{equation}
where $e$ is the particle charge and $A_\mu$ is the photon field, and implies the one- and two-photon vertices
\begin{eqnarray}
\langle p_f|V^{(1)\mu}_{em}|p_i\rangle&=&ie(p_f+p_i)^\mu,\nonumber\\
\langle p_f|V^{(2)\mu\nu}_{em}|p_i\rangle &=&2ie^2\eta^{\mu\nu}.
\end{eqnarray}

The corresponding charged massive spin-1 Lagrangian has the Proca form~\cite{bjd}
\begin{equation}
{\cal L}_0^{S=1}=-{1\over 2}\,B^\dagger_{\mu\nu}B^{\mu\nu}+m^2\,B^\dagger_\mu B^\mu\,,
\end{equation}
where $B^\mu$ is a spin one field subject to the constraint $\partial_\mu B^\mu=0$ and $B^{\mu\nu}$ is the antisymmetric tensor
\begin{equation}
B^{\mu\nu}=\partial^\mu B^\nu-\partial^\nu B^\mu\,.\label{eq:jv}
\end{equation}
The minimal substitution then leads to the interaction Lagrangian
\begin{equation}
{\cal L}_{int}^{S=1}=i\,e\,A^\mu B^{\nu\dagger}\left(\eta_{\nu\alpha}\overleftrightarrow{\partial}_\mu-\eta_{\alpha\mu}\overleftrightarrow{\partial}_\nu\right)B^\alpha
-e^2A^\mu A^\nu (\eta_{\mu\nu}\eta_{\alpha\beta}-\eta_{\mu\alpha}\eta_{\nu\beta})B^{\alpha\dagger}B^\beta\,,
\end{equation}
and the one, two photon vertices
\begin{eqnarray}
\big\langle p_f,\epsilon_B\,\big|V_{em}^{(1)\mu}\big|\,p_i,\epsilon_A\big\rangle_{S=1}&=&-i\,e\,\epsilon_{B\beta}^*\left((p_f+p_i)^\mu\eta^{\alpha\beta}
-\eta^{\beta\mu}p_f^\alpha-\eta^{\alpha\mu}p_i^\beta\right)\epsilon_{A\alpha}\,,\nonumber\\
\big\langle p_f,\epsilon_B\,\big|V_{em}^{(2)\mu\nu}\big|\,p_i,\epsilon_A\big\rangle_{S=1}&=&i\,e^2\,\epsilon_{B\beta}^*\left(2\eta^{\alpha\beta}\eta^{\mu\nu}
-\eta^{\alpha\mu}\eta^{\beta\nu}-\eta^{\alpha\nu}\eta^{\beta\mu}\right)\epsilon_{A\alpha}\,.\label{eq:kb}
\end{eqnarray}
However, Eq.~\eqref{eq:kb} is {\it not} the correct result for a fundamental spin-1 particle such as the charged $W$-boson.  Because the $W$ arises in a gauge theory, the field tensor is not given by Eq. (\ref{eq:jv}) but rather is generated from the charged---$\sqrt{1\over 2}(x\pm iy)$---component of
\begin{equation}
\boldsymbol{B}_{\mu\nu}=D_\mu \boldsymbol{B}_\nu-D_\nu\boldsymbol{B}_\mu-g_{ga}\boldsymbol{B}_\mu\times\boldsymbol{B}_\nu\,,
\end{equation}
where $g_{ga}$ is the gauge coupling.  This modification implies the existence of an additional $W^\pm\gamma$ interaction, leading to an ``extra" contribution to the single photon vertex
\begin{equation}
\big\langle p_f,\epsilon_B\,\big|\delta V_{em}^{(1)\mu}\big|\,p_i,\epsilon_A\rangle_{S=1}=i\,e\,\epsilon_{B\beta}^*\left(\eta^{\alpha\mu}(p_i-p_f)^\beta
-\eta^{\beta\mu}(p_i-p_f)^\alpha)\right)\epsilon_{A\alpha}\,.\label{eq:jb}
\end{equation}
The significance of this term can be seen by using the mass-shell Proca constraints $p_i\cdot\epsilon_A=p_f\cdot\epsilon_B=0$ to write the total on-shell single photon vertex as
\begin{eqnarray}
\big\langle p_f,\epsilon_B\,\big|(V_{em}+\delta V_{em})^\mu\big|\,p_i,\epsilon_A\big\rangle_{S=1}&=&-i\,e\,\epsilon_{B\beta}^*
\left((p_f+p_i)^\mu\eta^{\alpha\beta}-2\eta^{\beta\mu}(p_i-p_f)^\alpha\right.\nonumber\\
&-&\left.2\eta^{\alpha\mu}(p_i-p_f)^\beta\right)\epsilon_{A\alpha}\,,
\end{eqnarray}
wherein, comparing with Eq. \ref{eq:jb}, we observe that the coefficient of the term $-\eta^{\alpha\mu}(p_i-p_f)^\beta+\eta^{\beta\mu}(p_i-p_f)^\alpha$ has been modified from unity to two.  Since the rest frame spin operator can be identified via\footnote{Equivalently, one can use the relativistic identity
\begin{equation}
\epsilon^*_{B\mu}q\cdot\epsilon_A-\epsilon_{A\mu}q\cdot\epsilon_B^*={1\over 1-{q^2\over m^2}}\left({i\over m}\epsilon_{\mu\beta\gamma\delta}p_i^\beta q^\gamma S^\delta-{1\over 2m}(p_f+p_i)_\mu\epsilon_B^*\cdot q\epsilon_A\cdot q\right),
\end{equation}
where $S^\delta={i\over 2m}\epsilon^{\delta\sigma\tau\zeta}\epsilon^*_{B\sigma}\epsilon_{A\tau}(p_f+p_i)_\zeta$ is the spin four-vector.}
\begin{equation}
B^\dagger_i B_j-B^\dagger_j
B_i=-i\,\epsilon_{ijk}\big\langle f\,\big|S_k\big|\,i\big\rangle\label{eq:mm}\,,
\end{equation}
the corresponding piece of the nonrelativistic interaction Lagrangian becomes
\begin{equation}
{\cal L}_{\rm int}=-g\,{e\over 2m} \big\langle f\,\big|\boldsymbol{S}\big|\,i\big\rangle\cdot\boldsymbol{\nabla}\times\boldsymbol{A}\,,
\end{equation}
where $g$ is the gyromagnetic ratio and we have included a factor
$2m$ which accounts for the normalization condition of the spin one
field.  Thus the ``extra" interaction required by a gauge theory
changes the $g$-factor from its Belinfante value of unity~\cite{bfn}
to its universal value of two, as originally proposed by Weinberg~\cite{wbg} and
more recently buttressed by a number of additional arguments~\cite{lrg}.
Henceforth in this manuscript then we shall assume the $g$-factor of the
spin-1 system to have its ``natural'' value $g=2$, since it is in this
case that the high-energy properties of the scattering are well
controlled and the factorization properties of gravitational amplitudes are valid~\cite{hbp}.

\section{Compton Scattering}\label{sec:compton}

\begin{figure}[h]
\begin{center}
\includegraphics[width=12cm]{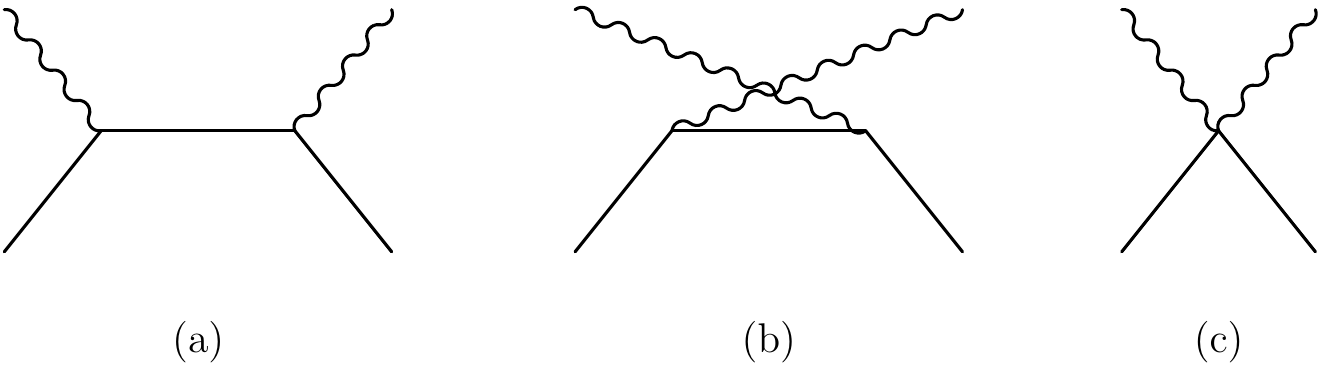}
\caption{Diagrams relevant to Compton scattering.}\label{fig:compton}
\end{center}
\end{figure}

The vertices given in the previous section can now be used to evaluate the ordinary Compton scattering amplitude,
$$\gamma+S\rightarrow\gamma+S,$$
for a spin-1 system having charge $e$ and mass $m$ by summing the contributions of the three diagrams shown in Figure~\ref{fig:compton}, yielding
\begin{eqnarray}
{\rm Amp}^{\rm Comp}_{S=1}
&=&2e^2\Bigg\{\epsilon_A\cdot\epsilon_B^*\left[{\epsilon_i\cdot
p_i\epsilon_f^*\cdot p_f\over p_i\cdot k_i}-{\epsilon_i\cdot
p_f\epsilon_f^*\cdot p_i\over p_i\cdot
k_f}-\epsilon_i\cdot\epsilon_f^*\right]\nonumber\\
&-&\left[\epsilon_A\cdot
[\epsilon_f^*,k_f]\cdot\epsilon_B^*\left({\epsilon_i\cdot p_i\over
p_i\cdot k_i} -{\epsilon_i\cdot p_f\over p_i\cdot k_f}\right)
- \epsilon_A\cdot[\epsilon_i,k_i]\cdot\epsilon_B^*\left({\epsilon_f\cdot
p_f\over p_i\cdot k_i}
-{\epsilon_f^*\cdot p_i\over p_i\cdot k_f}\right)\right]\nonumber\\
&-&\left[{1\over p_i\cdot
k_i}\epsilon_A\cdot[\epsilon_i,k_i]\cdot[\epsilon_f^*,k_f]\cdot\epsilon_B^*
-{1\over p_i\cdot
k_f}\epsilon_A\cdot[\epsilon_f^*,k_f]\cdot[\epsilon_i,k_i]\epsilon_B^*\right]\Bigg\},
\label{eq:gh}
\end{eqnarray}
with the momentum conservation condition $p_i+k_i=p_f+k_f$.
We can verify the gauge invariance of the above form by noting that this amplitude
can be written in the equivalent form
\begin{eqnarray}
{\rm Amp}_{S=1}^{\rm Comp}&=&{2e^2\over p_i\cdot k_ip_i\cdot k_f}\Bigg\{\epsilon_B^*\cdot\epsilon_A(p_i\cdot F_i\cdot F_f\cdot p_i)\nonumber\\
&+&\Big[(\epsilon_B^*\cdot F_f\cdot\epsilon_A)(p_i\cdot F_i\cdot p_f)+(\epsilon_B^*\cdot F_i\cdot\epsilon_A)(p_i\cdot F_f\cdot p_f)\Big]\nonumber\\
&-&\Big[p_i\cdot k_f(\epsilon_B^*\cdot F_f\cdot F_i\cdot \epsilon_A)-p_i\cdot k_i(\epsilon_B^*\cdot F_i\cdot F_f\cdot\epsilon_A)\Big]\Bigg\}\,,
\end{eqnarray}
where the electromagnetic field tensors are $F_i^{\mu\nu}=\epsilon_i^\mu k_i^\nu-\epsilon_i^\nu k_i^\mu$ and $F_f^{\mu\nu}=\epsilon_f^{*\mu} k_f^\nu-\epsilon_f^{*\nu} k_f^\mu$.  Since $F_{i,f}$ are obviously invariant under the substitutions $\epsilon_{i,f}\rightarrow \epsilon_{i,f}+\lambda k_{i,f},\,\,i=1,2$, it is clear that Eq.~\eqref{eq:gh} satisfies the gauge invariance strictures
\begin{equation}
\epsilon_f^{*\mu}k_i^\nu{\rm Amp}^{\rm Comp}_{\mu\nu,S=1}=k_f^\mu\epsilon_i^\nu{\rm Amp}^{\rm Comp}_{\mu\nu,S=1}=0\,.
\end{equation}
In order to make the transition to gravity, it is useful to utilize the helicity formalism~\cite{hct}, wherein one evaluates the matrix elements of the Compton amplitude between initial and final spin-1 and photon states having definite helicity, where helicity is defined as the projection of the particle spin along the momentum direction. We work initially in the center of mass frame and, for a photon incident with four-momentum $k_{i}^\mu=p_{\rm CM}(1,\hat{z})$, we choose the polarization vectors
\begin{equation}
\epsilon_i^{\lambda_i}=-{\lambda_i\over
\sqrt{2}}\big(\hat{x}+i\lambda_i \hat{y}\big),\qquad \lambda_i=\pm\,,
\end{equation}
while for an outgoing photon with $k_{f}^\mu=p_{\rm CM}(1,\cos\theta_{\rm CM}\hat{z}+\sin\theta_{\rm CM}\hat{x})$ we use polarizations
\begin{equation}
\epsilon_f^{\lambda_f}=-{\lambda_f\over
\sqrt{2}}\Big(\cos\theta_{\rm CM}\hat{x}+
i\lambda_f\hat{y}-\sin\theta_{\rm CM}\hat{z}\Big),\quad \lambda_f=\pm\,.
\end{equation}
We can define corresponding helicity states for the spin-1 system.  In this case the initial and final four-momenta are
$p_i^\mu=(E_{\rm CM},-p_{\rm CM}\hat{z})$ and $p_f^\mu=\Big(E_{\rm CM},-p_{\rm CM}(\cos\theta_{\rm CM}\hat{z}+\sin\theta_{\rm CM}\hat{x})\Big)$ and there exist two transverse polarization four-vectors
\begin{eqnarray}
\epsilon_{A}^{\pm\mu}&=&\Bigg(0,{\pm\hat{x}-i\hat{y}\over \sqrt2}\Bigg)\,,\nonumber\\
\epsilon_{B}^{\pm\mu}&=&\Bigg(0,{\pm\cos\theta_{\rm CM}\hat{x}+ i\hat{y}\mp\sin\theta_{\rm CM}\hat{z}\over \sqrt 2}\Bigg)\,,
\end{eqnarray}
in addition to the longitudinal mode with polarization four-vectors
\begin{eqnarray}
\epsilon_{A}^{0\mu}&=&{1\over m}\big(p_{\rm CM},-E_{\rm CM}\hat{z}\big)\,,\nonumber\\
\epsilon_{B}^{0\mu}&=&{1\over m}\Big(p_{\rm CM},-E_{\rm CM}(\cos\theta_{\rm CM}\hat{z}+\sin\theta_{\rm CM}\hat{x})\Big)\,,
\end{eqnarray}
In terms of the usual invariant kinematic (Mandelstam) variables
\begin{equation}
s=\big(p_i+k_i\big)^2,\quad t=\big(k_i-k_f\big)^2,\quad u=\big(p_i-k_f\big)^2\,,\end{equation}
we identify
\begin{eqnarray}
p_{\rm CM}&=&{s-m^2\over 2\sqrt{s}}\,,\nonumber\\
E_{\rm CM}&=&{s+m^2\over 2\sqrt{s}}\,,\nonumber\\
\cos{1\over
2}\theta_{\rm CM}&=&{\Big((s-m^2)^2+s t\Big)^{1\over 2}\over s-m^2}={\Big(m^4-s u\Big)^{1\over 2}\over s-m^2}\,,\nonumber\\
\sin{1\over 2}\theta_{\rm CM}&=&{\big(-s t\big)^{1\over 2}\over s-m^2}\,.\label{eq:bv0}
\end{eqnarray}
The invariant cross-section for unpolarized Compton scattering is then given by
\begin{equation}
{d\sigma_{S=1}^{\rm Comp}\over dt}={1\over 16\pi(s-m^2)^2}\ {1\over
3}\sum_{a,b=-,0,+}{1\over 2}\sum_{c,d=-,+}\Big|B^1(ab;cd)\Big|^2\,,\label{eq:nh}\displaystyle
\end{equation}
where
\begin{equation}
B^1(ab;cd)=\big \langle p_f,b;k_f,d\,\big | {\rm Amp}_{S=1}^{\rm Comp}\big | \,p_i,a;k_i,c\,\big\rangle\,,
\end{equation}
is the Compton amplitude for scattering of a photon with four-momentum $k_i$, helicity a from a spin-1 target having four-momentum $p_i$, helicity $c$ to a photon with four-momentum $k_f$, helicity $d$ and target with four-momentum $p_f$, helicity $b$.  The helicity amplitudes can now be calculated straightforwardly.  There exist $3^2\times2^2=36$ such amplitudes but, since helicity reverses under spatial inversion, parity invariance of the electromagnetic interaction requires that\footnote{Note that we require only that the magnitudes of the helicity amplitudes related by parity and/or time reversal be the same.  There could exist unobservable phases.}
$$\big|B^1(ab;cd)\big|=\big|B^1(-a-b;-c-d)\big|\,.$$
Also, since helicity is unchanged under time reversal, but initial and final states are interchanged, T-invariance of the electromagnetic interaction requires that
$$\big|B^1(ab;cd)\big|=\big|B^1(ba;dc)\big|\,.$$
Consequently there exist only {\it twelve} independent helicity amplitudes.  Using Eq.~\eqref{eq:gh} we calculate the various helicity amplitudes in the center of mass frame and then write these results in terms of invariants using Eq.~\eqref{eq:bv0}, yielding
\begin{eqnarray}
\big |B^1(++;++)\big |&=&\big |B^1(--;--)\big |=2e^2{\big((s-m^2)^2+m^2t\big)^2\over (s-m^2)^3(u-m^2)}\,,\nonumber\\
\big |B^1(++;--)\big |&=&\big |B^1(--;++)\big |=2e^2{(m^4-su)^2\over (s-m^2)^3 (u-m^2)}\,,\nonumber\\
\big |B^1(+-;+-)\big |&=&\big |B^1(-+;-+)\big |=2e^2{m^4t^2\over (s-m^2)^3(u-m^2)}\,,\nonumber\\
\big |B^1(+-;-+)\big |&=&\big |B^1(-+;+-)\big |=2e^2{s^2t^2\over(s-m^2)^3(u-m^2)}\,,\nonumber\\
\big |B^1(++;+-)\big |&=&\big |B^1(--;-+)\big |=\big|B^1(++;-+)\big|=\big|B^1(--;+-)\big|\,,\nonumber\\
&=&2e^2{m^2t(m^4-su)\over(s-m^2)^3(u-m^2)}\,,\nonumber\\
\big|B^1(+-;++)\big|&=&\big|B^1(-+;--)\big|=\big|B^1(-+;++)\big|=\big|B^1(+-;--)\big|\,,\nonumber\\
&=&2e^2{m^2t(m^4-su)\over (s-m^2)^3(u-m^2)}\,.
\end{eqnarray}
and
\begin{eqnarray}
\big|B^1(0+;++)\big|&=&\big|B^1(0-;--)\big|=\big|B^1(+0;++)\big|=\big|B^1(-0;--)\big|\,,\nonumber\\
&=&2e^2{\sqrt{2}m\big(t m^2+(s-m^2)^2\big)\sqrt{-t(m^4-su)}\over (s-m^2)^3(u-m^2)}\,,\nonumber\\
\big|B^1(0+;+-)\big|&=&\big|B^1(0-;-+)\big|=\big|B^1(+0;-+)\big|=\big|B^1(-0;+-)\big|\,,\nonumber\\
&=&2e^2{\sqrt{2}mst\sqrt{-t(m^4-su)}\over (s-m^2)^3(u-m^2)}\,,\nonumber\\
\big|B^1(0+;-+)\big|&=&\big|B^1(0-;+-)\big|=\big|B^1(+0;+-)\big|=\big|B^1(-0;-+)\big|\,,\nonumber\\
&=&2e^2{\sqrt{2}m^3t\sqrt{-t(m^4-su)}\over (s-m^2)^3(u-m^2)}\,,\nonumber\\
\big|B^1(0+;--)\big|&=&\big|B^1(0-;++)\big|=\big|B^1(+0;--)\big|=\big|B^1(-0;++)\big|\,,\nonumber\\
&=&2e^2{\sqrt{2}m\big(-t(m^4-su)\big)^{3\over 2}\over (s-m^2)^3t(u-m^2)}\,,\nonumber\\
\big|B^1(00;++)\big|&=&\big|B^1(00;--)\big|=2e^2{\big(2tm^2+(s-m^2)^2\big)(m^4-su)\over (s-m^2)^3(u-m^2)}\,,\nonumber\\
\big|B^1(00;+-)\big|&=&\big|B^1(00;-+)\big|=2e^2{\big(m^2t((s-m^2)^2+2st\big)\over (s-m^2)^3(u-m^2)}\,.
\end{eqnarray}
Substitution into Eq.~\eqref{eq:nh} then yields the invariant cross-section for unpolarized Compton scattering from a charged spin-1 target
\begin{equation}
{d\sigma^{\rm Comp}_{S=1}\over dt}={e^4\over 12\pi(s-m^2)^4(u-m^2)^2}\Big[(m^4-su+t^2)\big(3(m^4-su)+t^2\big)+t^2(t-m^2)(t-3m^2)\Big]\,,
\end{equation}
which can be compared with the corresponding results for unpolarized Compton scattering from charged spin-0 and spin-$\frac12$ targets found in ref.~\cite{ppa}---
\begin{eqnarray}
{d\sigma_{S=0}^{\rm Comp}\over dt}&=&{e^4\over 4\pi(s-m^2)^4(u-m^2)^2}\left[(m^4-su)^2+m^4t^2\right]\,,\nonumber\\
{d\sigma_{S={1\over 2}}^{\rm Comp}\over dt}&=&{e^4\over 8\pi{(s-m^2)^4(u-m^2)^2}}\Big[(m^4-su)\big(2(m^4-su)+t^2\big)+m^2t^2(2m^2-t)\Big]\,.\nonumber\\
\quad
\end{eqnarray}
Often such results are written in the {\it laboratory} frame, wherein the target is at rest, by use of the relations
\begin{eqnarray}
s-m^2&=&2m\omega_i,\quad u-m^2=-2m\omega_f\,,\nonumber\\[1ex]
m^4-su&=&4m^2\omega_i\omega_f\cos^2{\theta_L\over 2},\quad
m^2t=-4m^2\omega_i\omega_f\sin^2{\theta_L\over 2}\,,
\end{eqnarray}
and
\begin{equation}
{dt\over d\Omega}={d\over 2\pi
d\cos\theta_L}\left(-{2\omega_i^2(1-\cos\theta_L)\over
1+{\omega_i\over m}(1-\cos\theta_L)}\right)={\omega_f^2\over \pi}\,.
\end{equation}
Introducing the fine structure constant $\alpha=e^2/4\pi$, we find then
\begin{eqnarray}
{d\sigma_{{\rm lab},S=1}^{\rm Comp}\over d\Omega}&=&{\alpha^2\over m^2}{\omega_f^4\over \omega_i^4}\left[\bigg(\cos^4{\theta_L\over 2}+\sin^4{\theta_L\over 2}\bigg)\bigg(1+2{\omega_i\over m}\sin^2{\theta_L\over 2}\bigg)^2\right.\nonumber\\
&+&\left.{16\omega_i^2\over 3m^2}\sin^4{\theta_L\over 2}\bigg(1+2{\omega_i\over m}\sin^2{\theta_L\over 2}\bigg)+{32\omega_i^4\over 3m^4}\sin^8{\theta_L\over 2}\right]\,,\nonumber\\
{d\sigma^{\rm Comp}_{{\rm lab},S={1\over 2}}\over d\Omega}&=&{\alpha^2\over m^2}{\omega_f^3\over \omega_i^3}\left[\bigg(\cos^4{\theta_L\over 2}+\sin^4{\theta_L\over 2}\bigg)\bigg(1+2{\omega_i\over m}\sin^2{\theta_L\over 2}\bigg)+2{\omega_i^2\over m^2}\sin^4{\theta_L\over 2}\right]\,,\nonumber\\
{d\sigma^{\rm Comp}_{{\rm lab},S=0}\over d\Omega}&=&{\alpha^2\over m^2}{\omega_f^2\over\omega_i^2}\left[\cos^4{\theta_L\over 2}+\sin^4{\theta_L\over 2}\right]\,.
\end{eqnarray}
We observe that the nonrelativistic laboratory cross-section has an identical form for {\it any} spin
\begin{equation}\label{e:NRphoto}
\left.{d\sigma_{{\rm lab},S}^{\rm Comp}\over d\Omega}\right|^{NR}={\alpha^2\over m^2}\left[\bigg(\cos^4{\theta_L\over 2}+\sin^4{\theta_L\over 2}\Big)\Big(1+{\cal O}\Big({\omega_i\over m}\Big)\bigg)\right]\,,
\end{equation}
which follows from the universal form of the Compton amplitude for scattering from a spin-$S$ target in the low-energy ($\omega\ll m$) limit, which in turn arises from the universal form of the Compton amplitude for scattering from a spin-$S$ target in the low-energy limit---
\begin{equation}
\big\langle S,M_f;\epsilon_f\,\big|{\rm Amp}^{\rm Comp}_S\big|\,S,M_i;\epsilon_i\big\rangle_{\omega\ll m}=2e^2\,\epsilon_f^*\cdot\epsilon_i\,\delta_{M_i,M_f}+\ldots\,,
\end{equation}
which obtains in an effective field theory approach to Compton
scattering~\cite{wsb}.\footnote{ That the seagull contribution
dominates the non relativistic cross-section is clear from the feature that
\begin{equation}
{\rm Amp}_{\rm Born}\sim 2e^2{\epsilon_f^*\cdot p\epsilon_i\cdot p\over p\cdot k}\sim{\omega\over m}\times{\rm Amp}_{\rm seagull}=2e^2\epsilon_f^*\cdot\epsilon_i.
\end{equation}}

\section{Gravitational Interactions}\label{sec:grav}

In the previous section we discussed the treatment the familiar electromagnetic interaction, using Compton scattering on a spin-1 target as an example.  In this section we show how the gravitational interaction can be evaluated via methods parallel to those used in the electromagnetic case.  An important difference is that while in the electromagnetic case we have the simple interaction Lagrangian
\begin{equation}
{\cal L}_{int}=-eA_\mu J^\mu\,,
\end{equation}
where $J^\mu$ is the electromagnetic current matrix element, for gravity we have
\begin{equation}
{\cal L}_{int}={\kappa\over 2}h_{\mu\nu}T^{\mu\nu}\,.
\end{equation}
Here the field tensor $h_{\mu\nu}$ is defined in terms of the metric via
\begin{equation}
g_{\mu\nu}=\eta_{\mu\nu}+\kappa h_{\mu\nu}\,,
\end{equation}
where $\kappa$ is given in terms of the Cavendish constant $G$ by
$\kappa^2=32\pi G$. The Einstein-Hilbert action is
\begin{equation}
  \label{e:EH}
  \mathcal S_{\rm Einstein-Hilbert}=\int d^4x \,\sqrt{-g} \, {2\over
    \kappa^2}\, R\,,
\end{equation}
where
\begin{equation}
\sqrt{-g}\equiv\sqrt{-{\rm det}\,g}=\exp{1\over 2}{\rm tr}{\rm log}\, g=1+{1\over 2}\eta^{\mu\nu}h_{\mu\nu}+\ldots\,,
\end{equation}
is the square root of the determinant of the metric and $R=
R^\lambda{}_{\mu\lambda\nu} g^{\mu\nu}$ is the Ricci scalar curvature obtained by contracting the Riemann tensor
$R^\mu{}_{\nu\rho\sigma}$ with the metric tensor.  The energy-momentum tensor is defined in
terms of the matter Lagrangian via
\begin{equation}
T_{\mu\nu}={2\over \sqrt{-g}}{\delta {\sqrt{-g}\cal L}_{\rm mat}\over
\delta g^{\mu\nu}}\,.\label{eq:pk}
\end{equation}
\begin{figure}[h]
\begin{center}
\includegraphics[height=4cm,width=10cm]{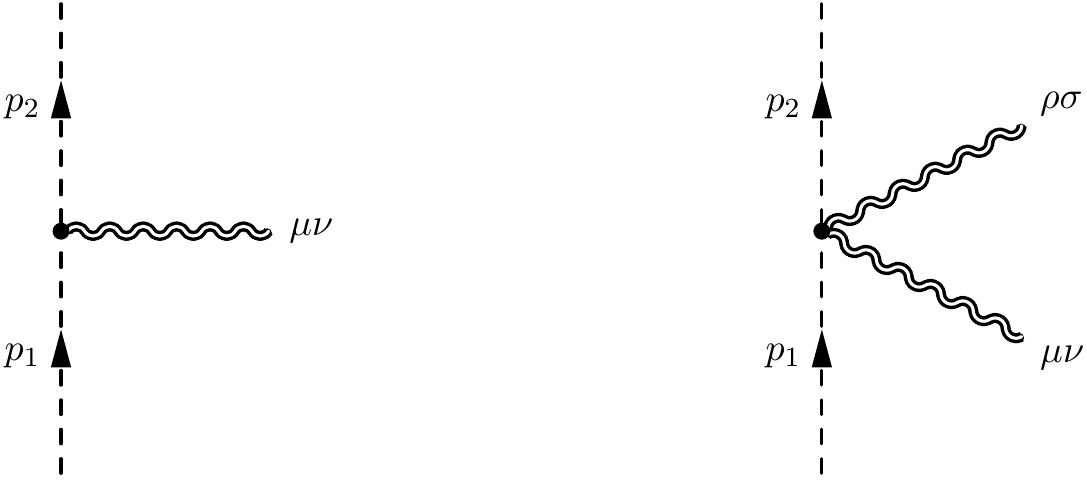}
\caption{ (a) The one-graviton and (b) two-graviton emission vertices from either a
  scalar, spinor or vector particle. }\label{fig:vertices}
\end{center}
\end{figure}
The spin-1 single graviton emission vertex shown in figure~\ref{fig:vertices}(a) can now be identified
\begin{eqnarray}
\big\langle p_f,\epsilon_B\,\big|V_{grav}^{(1)\mu\nu}\big|\,p_i,\epsilon_A\big\rangle_{S=1}&=&-i\,{\kappa\over 2}\,\bigg[\epsilon_B^*\cdot\epsilon_A\big(p_i^\mu p_f^\nu+p_i^\nu p_f^\mu\big)-\epsilon_B^*\cdot p_i\,\big(p_f^\mu\epsilon_A^\nu+\epsilon_A^\mu p_f^\nu\big)\nonumber\\
& -&\epsilon_A\cdot p_f\big(p_i^\nu\epsilon_B^{*\mu}+p_i^\mu\epsilon_B^{*\nu}\big)+\big(p_f\cdot p_i-m^2\big)\big(\epsilon_A^\mu\epsilon_B^{*\nu}+\epsilon_A^\nu\epsilon_B^{*\mu}\big)\nonumber\\
& -&\eta^{\mu\nu}\Big[\big(p_i\cdot p_f-m^2\big)\epsilon_B^*\cdot\epsilon_A-\epsilon_B^*\cdot p_i\,\epsilon_A\cdot p_f\Big]\bigg]\,.\label{eq:dt}
\end{eqnarray}
There also exist two-graviton (seagull) vertices shown in figure~\ref{fig:vertices}(b), which
can be found by expanding the stress-energy tensor to second order in $h_{\mu\nu}$.
\begin{eqnarray}
&&\big\langle p_f, \epsilon_B;k_f\,\big|V_{grav}^{(2)\,\mu\nu,\rho\sigma}\big|\,p_i,\epsilon_A;k_i\big\rangle_{S=1}\!\!\!\!=-i\,{\kappa^2\over
4}\epsilon_A^\alpha(\epsilon_B^\beta)^*\,\Bigg\{\nonumber\\ \nonumber&+&\Big[p_{i\beta}p_{f\alpha}
-\left.
          \eta_{\alpha\beta}(p_i\cdot p_f - m^2)\Big]
      \Big(\eta_{\mu\rho}\eta_{\nu\sigma}+
          \eta_{\mu\sigma}\eta_{\nu\rho} -
          \eta_{\mu\nu}\eta_{\rho\sigma}\Big)\right.\nonumber\\
          &+&\left.
    \eta_{\mu\rho}\Big[\eta_{\alpha\beta}\Big(p_{i\nu}p_{f\sigma} +
                p_{i\sigma}p_{f\nu}\Big) -
          \eta_{\alpha\nu}p_{i\beta}p_{f\sigma}-
          \eta_{\beta\nu}p_{i\sigma}p_{f\alpha}\right.\nonumber\\
          &-&\left.
          \eta_{\beta\sigma}p_{i\nu}p_{f\alpha} -
          \eta_{\alpha\sigma}p_{i\beta}
            p_{f\nu} + (p_i\cdot p_f -
                m^2\Big)\Big(\eta_{\alpha\nu}\eta_{\beta\sigma} +
                \eta_{\alpha\sigma}\eta_{\beta\nu}\Big)\Big]\right.\nonumber\\
                &+&\left.
    \eta_{\mu\sigma}\Big[\eta_{\alpha\beta}\Big(p_{i\nu}p_{f\rho} +
                p_{i\rho}p_{f\nu}\Big) -
          \eta_{\alpha\nu}p_{i\beta}p_{f\rho} -
          \eta_{\beta\nu}p_{i\rho}p_{f\alpha}\right.\nonumber\\
          &-&\left.
          \eta_{\beta\rho}p_{i\nu}p_{f\alpha}-
          \eta_{\alpha\rho}p_{i\beta}
            p_{f\nu} + (p_i\cdot p_f -
                m^2\Big)\eta_{\alpha\nu}\eta_{\beta\rho} +
                \eta_{\alpha\rho}\eta_{\beta\nu}\Big)\Big]\right.\nonumber\\
                &+&\left.
    \eta_{\nu\rho}\Big[\eta_{\alpha\beta}\Big(p_{i\mu}p_{f\sigma} +
                p_{i\sigma}p_{f\mu}\Big)
                -\eta_{\alpha\mu}p_{i\beta}p_{f\sigma} -
          \eta_{\beta\mu}p_{i\sigma}p_{f\alpha}\right.\nonumber\\
          &-&\left.\eta_{\beta\sigma}p_{i\mu}p_{f\alpha}
          -\eta_{\alpha\sigma}p_{i\beta}
            p_{f\mu} + (p_i\cdot p_f -
                m^2\Big)\Big(\eta_{\alpha\mu}\eta_{\beta\sigma} +
                \eta_{\alpha\sigma}\eta_{\beta\mu}\Big)\Big]\right.\nonumber\\
                &+&\left.
    \eta_{\nu\sigma}\Big[\eta_{\alpha\beta}\Big(p_{i\mu}p_{f\rho} +
                p_{i\rho}p_{f\mu}\Big) -
          \eta_{\alpha\mu}p_{i\beta}p_{f\rho} -
          \eta_{\beta\mu}p_{i\rho}p_{f\alpha}\right.\nonumber\\
          &-&\left.\eta_{\beta\rho}p_{i\mu}p_{f\alpha}-\eta_{\alpha\rho}p_{i\beta}
            p_{f\mu} + (p_i\cdot p_f -
                m^2\Big)\Big(\eta_{\alpha\mu}\eta_{\beta\rho} +
                \eta_{\alpha\rho}\eta_{\beta\mu}\Big)\Big]\right.\nonumber\\
                &-&\left.
    \eta_{\mu\nu}\Big[\eta_{\alpha\beta}\Big(p_{i\rho}p_{f\sigma} +
                p_{i\sigma}p_{f\rho}\Big) -
          \eta_{\alpha\rho}p_{i\beta}p_{f\sigma} -
          \eta_{\beta\rho}p_{i\sigma}p_{f\alpha}\right.\nonumber\\
          &-&\left.\eta_{\beta\sigma}p_{i\rho}p_{f\alpha}-
          \eta_{\alpha\sigma}p_{i\beta}p_{f\rho} + \Big(p_i\cdot p_f -
                m^2\Big)\Big(\eta_{\alpha\rho}\eta_{\beta\sigma} +
                \eta_{\beta\rho}\eta_{\alpha\sigma}\Big)\Big]\right.\nonumber\\
                &-&\left.
    \eta_{\rho\sigma}\Big[\eta_{\alpha\beta}\Big(p_{i\mu}p_{f\nu} +
                p_{i\nu}p_{f\mu}\Big) -
          \eta_{\alpha\mu}p_{i\beta}p_{f\nu} -
          \eta_{\beta\mu}p_{i\nu}p_{f\alpha}\right.\nonumber\\
          &-&\left.
          \eta_{\beta\nu}p_{i\mu}p_{f\alpha} -
          \eta_{\alpha\nu}p_{i\beta}
            p_{f\mu} + (p_i\cdot p_f -
                m^2\Big)\Big(\eta_{\alpha\mu}\eta_{\beta\nu} +
                \eta_{\beta\mu}\eta_{\alpha\nu}\Big)\Big]\right.\nonumber\\
                 &+&\left.
    \Big(\eta_{\alpha\rho}p_{i\mu} -
          \eta_{\alpha\mu}p_{i\rho}\Big)\Big(\eta_{\beta\sigma}
            p_{f\nu} - \eta_{\beta\mu}p_{f\sigma}\Big)\right.\nonumber\\
            &+&\left.
    \Big(\eta_{\alpha\sigma}p_{i\nu} -
          \eta_{\alpha\nu}p_{i\sigma}\Big)\Big(\eta_{\beta\rho}
            p_{f\mu} - \eta_{\beta\mu}p_{f\rho}\Big)\right.\nonumber\\
            &+&\left.
    \Big(\eta_{\alpha\sigma}p_{i\mu} -
          \eta_{\alpha\mu}p_{i\sigma}\Big)\Big(\eta_{\beta\rho}
          p_{f\nu} - \eta_{\beta\nu}p_{f\rho}\Big)\right.\nonumber\\
          &+&
    \Big(\eta_{\alpha\rho}p_{i\nu} -
          \eta_{\alpha\nu}p_{i\rho}\Big)\Big(\eta_{\beta\sigma}
            p_{f\mu} - \eta_{\beta\mu}p_{f\sigma}\Big) \Bigg\}\,.
\end{eqnarray}
\begin{figure}[h]
\begin{center}
\includegraphics[width=5cm]{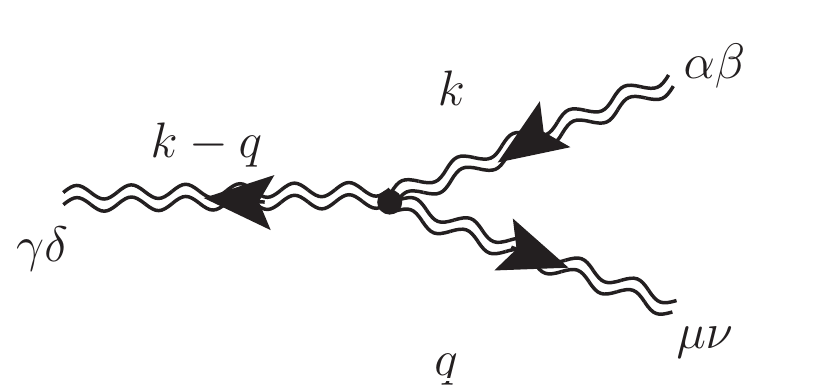}
 \caption{ The three
  graviton vertex }\label{fig:3grav}
\end{center}
\end{figure}
Finally, we require the triple graviton vertex of figure~\ref{fig:3grav}
\begin{eqnarray}
\tau^{\mu\nu}_{\alpha\beta,\gamma\delta}(k,q)\!\!\!\!&=&\!\!\!-{i\,\kappa\over
2}\left[ \big(I_{\alpha\beta,\gamma\delta}-{1\over
2}\eta_{\alpha\beta}\eta_{\gamma\delta}\big)\left.\bigg[k^\mu
k^\nu+(k-q)^\mu (k-q)^\nu+q^\mu q^\nu-{3\over
2}\eta^{\mu\nu}q^2\right]\right.\nonumber\\
&+&\left.2q_\lambda
q_\sigma\left.\Big[I^{\lambda\sigma,}{}_{\alpha\beta}I^{\mu\nu,}
{}_{\gamma\delta}+I^{\lambda\sigma,}{}_{\gamma\delta}I^{\mu\nu,}
{}_{\alpha\beta}-I^{\lambda\mu,}{}_{\alpha\beta}I^{\sigma\nu,}
{}_{\gamma\delta}-I^{\sigma\nu,}{}_{\alpha\beta}I^{\lambda\mu,}
{}_{\gamma\delta}\right]\right.\nonumber\\
&+&\left.\Big[q_\lambda
q^\mu\big(\eta_{\alpha\beta}I^{\lambda\nu,}{}_{\gamma\delta}
+\eta_{\gamma\delta}I^{\lambda\nu,}{}_{\alpha\beta})+ q_\lambda
q^\nu(\eta_{\alpha\beta}I^{\lambda\mu,}{}_{\gamma\delta}
+\eta_{\gamma\delta}I^{\lambda\mu,}{}_{\alpha\beta}\big)\right.\nonumber\\
&-&\left.q^2\big(\eta_{\alpha\beta}I^{\mu\nu,}{}_{\gamma\delta}+\eta_{\gamma\delta}
I^{\mu\nu,}{}_{\alpha\beta})-\eta^{\mu\nu}q^\lambda
q^\sigma(\eta_{\alpha\beta}
I_{\gamma\delta,\lambda\sigma}+\eta_{\gamma\delta}
I_{\alpha\beta,\lambda\sigma}\big)\Big]\right.\nonumber\\
&+&\left.\Big[2q^\lambda\big(I^{\sigma\nu,}{}_{\gamma\delta}
I_{\alpha\beta,\lambda\sigma}(k-q)^\mu
\!+\!I^{\sigma\mu,}{}_{\gamma\delta}I_{\alpha\beta,\lambda\sigma}(k-q)^\nu\right.\!-\!\left.I^{\sigma\nu,}{}_{\alpha\beta}I_{\gamma\delta,\lambda\sigma}k^\mu\!-\!
I^{\sigma\mu,}{}_{\alpha\beta}I_{\gamma\delta,\lambda\sigma}k^\nu\big)\right.\nonumber\\
&+&\left.q^2\big(I^{\sigma\mu,}{}_{\alpha\beta}I_{\gamma\delta,\sigma}{}^\nu+
I_{\alpha\beta,\sigma}{}^\nu
I^{\sigma\mu,}{}_{\gamma\delta}\big)+\eta^{\mu\nu}q^\lambda q_\sigma
\big(I_{\alpha\beta,\lambda\rho}I^{\rho\sigma,}{}_{\gamma\delta}+
I_{\gamma\delta,\lambda\rho}I^{\rho\sigma,}{}_{\alpha\beta}\big)\Big]\right.\nonumber\\
&+&\left.\Big[\big(k^2+(k-q)^2\big)\left(I^{\sigma\mu,}{}_{\alpha\beta}I_{\gamma\delta,\sigma}{}^\nu
+I^{\sigma\nu,}{}_{\alpha\beta}I_{\gamma\delta,\sigma}{}^\mu-{1\over
2}\eta^{\mu\nu}\big(I_{\alpha\beta,\gamma\delta}-{1\over 2}\eta_{\alpha\beta}\eta_{\gamma\delta})\right.\big)\right.\nonumber\\
&-&\left.\big(k^2\eta_{\alpha\beta}I^{\mu\nu,}{}_{\gamma\delta}+(k-q)^2\eta_{\gamma\delta}
I^{\mu\nu,}{}_{\alpha\beta}\big)\Big]\right.\bigg]\,,
\end{eqnarray}
where
\begin{equation}I_{\alpha\beta,\gamma\delta}={1\over 2}(\eta_{\alpha\gamma}\eta_{\beta\delta}+\eta_{\alpha\delta}\eta_{\beta\gamma})\,.\end{equation}

We work in harmonic (de Donder) gauge which satisfies, in
lowest order,
\begin{equation}
\partial^\mu h_{\mu\nu}={1\over 2}\partial_\nu h\,,\label{eq:dd}
\end{equation}
with
\begin{equation}
h\equiv{\rm tr}\,h_{\mu\nu}\,,
\end{equation}
in which the graviton propagator has the form
\begin{equation}
D_{\alpha\beta;\gamma\delta}(q)={i\over
q^2+i\epsilon}{1\over 2}(\eta_{\alpha\gamma}\eta_{\beta\delta}+\eta_{\alpha\delta}\eta_{\beta\gamma}
-\eta_{\alpha\beta}\eta_{\gamma\delta})\,.
\end{equation}
Then just as the (massless) photon is described in terms of a
spin-1 polarization vector $\epsilon_\mu$ which can have
projection (helicity) either plus- or minus-1 along the momentum
direction, the (massless) graviton is a spin-2 particle which
can have the projection (helicity) either plus- or minus-2 along
the momentum direction.  Since $h_{\mu\nu}$ is a symmetric tensor,
it can be described in terms of a direct product of unit spin
polarization vectors---
\begin{eqnarray}
{\rm helicity}&=&+2:\quad
h^{(2)}_{\mu\nu}=\epsilon^{+}_\mu \epsilon^{+}_\nu\,,\nonumber\\
{\rm helicity}&=&-2:\quad h^{(-2)}_{\mu\nu}=\epsilon^{-}_\mu
\epsilon^{-}_\nu\,,\label{eq:he}
\end{eqnarray}
and, just as in electromagnetism, there is a gauge condition---in
this case Eq.~\eqref{eq:dd}---which must be satisfied. Note that the
helicity states given in Eq.~\eqref{eq:he} are consistent with the
gauge requirement since
\begin{equation}
\eta^{\mu\nu}\epsilon_\mu^+\epsilon_\nu^+=\eta^{\mu\nu}\epsilon_\mu^-\epsilon_\nu^-=0,\quad{\rm and} \quad
k^\mu\epsilon_\mu^\pm=0\,.
\end{equation}
With this background we can now examine specific graviton reactions.

\subsection{Graviton Photo-production}\label{sec:gravphotoprod}

\begin{figure}[h]
\begin{center}
\includegraphics[width=10cm]{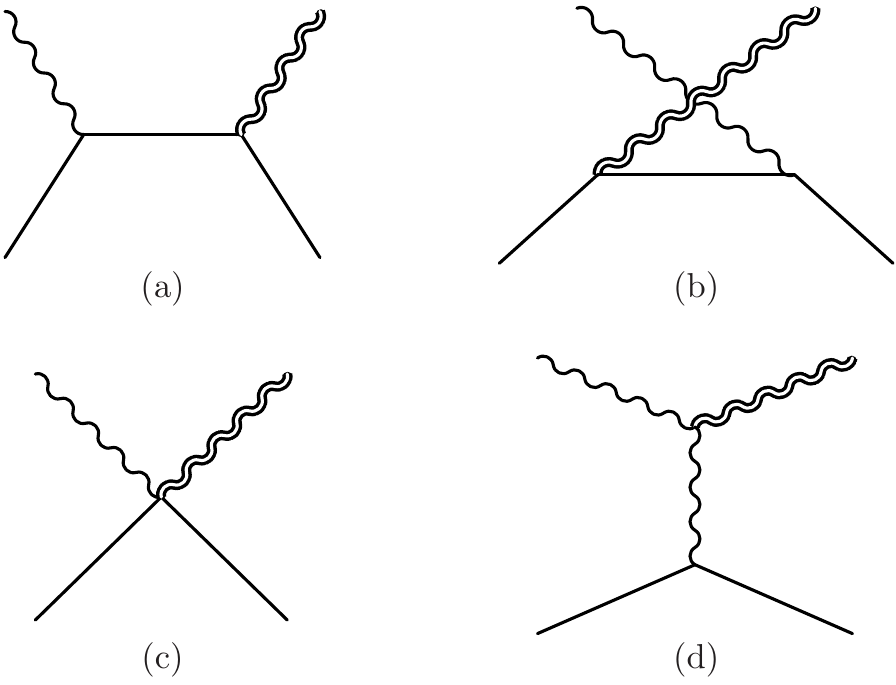}
 \caption{Diagrams relevant to graviton photo-production. }\label{fig:gravphoto}
\end{center}
\end{figure}

We first use the above results to discuss the problem of graviton photo-production on a spin-1 target---$\gamma+S\rightarrow g+S$---for which the relevant four diagrams are shown in Figure~\ref{fig:gravphoto}.  The electromagnetic and gravitational vertices needed for the Born terms and photon pole diagrams---Figures~\ref{fig:gravphoto}a,~\ref{fig:gravphoto}b, and~\ref{fig:gravphoto}d---have been given above.  For the photon pole diagram we require the graviton-photon coupling, which can be found from the electromagnetic energy-momentum tensor~\cite{jdj}
\begin{equation}
T_{\mu\nu}=-F_{\mu\alpha}F^\alpha_\nu+{1\over
4}g_{\mu\nu}F_{\alpha\beta}F^{\alpha\beta}\,,
\end{equation}
and yields the photon-graviton vertex\footnote{Note that this form agrees with the previously derived form for the massive graviton-spin-1 energy-momentum tensor---Eq.~\eqref{eq:dt}---in the $m\rightarrow 0$ limit.}
\begin{eqnarray}
\big\langle k_f,\epsilon_f\,\big|V_{grav}^{(\gamma)\mu\nu}\big|\,k_i,\epsilon_i\big\rangle&=&
i\,{\kappa\over 2}\left.\Big[\epsilon_f^*\cdot\epsilon_i\big(k_i^\mu k_f^\nu+k_i^\nu k_f^\mu\big)-\epsilon_f^*\cdot k_i\big(k_f^\mu\epsilon_i^\nu
+\epsilon_i^\mu k_f^\nu\big)\right.\nonumber\\
&-&\left.\epsilon_i\cdot k_f\big(k_i^\nu\epsilon_f^{*\mu}+k_i^\mu\epsilon_f^{*\nu}\big)+k_f\cdot k_i\big(\epsilon_i^\mu\epsilon_f^{*\nu}+\epsilon_i^\nu\epsilon_f^{*\mu}\big)\right.\nonumber\\
&-&\left.\eta^{\mu\nu}\left[k_f\cdot k_i\epsilon_f^*\cdot\epsilon_i-\epsilon_f^*\cdot k_i\epsilon_i\cdot k_f\right]\right.\Big]\,.\label{eq:gs}
\end{eqnarray}

Finally, we need the seagull vertex which arises from the feature that the energy-momentum tensor depends on $p_i,p_f$ and therefore yields a contact interaction when the minimal substitution is made, yielding the spin-1 seagull amplitude shown in Figure~\ref{fig:gravphoto}c.
\begin{eqnarray}
&&\big\langle p_f,\epsilon_B;k_f,\epsilon_f\epsilon_f\,\big|T\big|\,p_i,\epsilon_A;k_i,\epsilon_i\big\rangle_{\rm seagull}=\,{i\over 2}\,\kappa \,e\left.\Big[\epsilon_f^*\cdot(p_f+p_i)\,\epsilon_f^*\cdot\epsilon_i\,\epsilon_B^*\cdot\epsilon_A\,\right.\nonumber\\
&-&\left.\epsilon_B^*\cdot\epsilon_i\,\epsilon_f^*\cdot p_f\,\epsilon_f^*\cdot\epsilon_A-\epsilon_B^*\cdot p_i\,\epsilon_f^*\cdot\epsilon_i\,\epsilon_f^*\cdot\epsilon_A-\epsilon_A\cdot \epsilon_i\,\epsilon_f^*\cdot p_i\,\epsilon_f^*\cdot\epsilon_B^*\right.\nonumber\\
&-&\left.\epsilon_A\cdot p_f\,\epsilon_f^*\cdot\epsilon_i\,\epsilon_f^*\cdot\epsilon_B^*-\epsilon_f^*\cdot\epsilon_A\,\epsilon_i\cdot(p_f+p_i)\,\epsilon_f^*\cdot\epsilon_B^*\right.\Big]\,.
\end{eqnarray}
The individual contributions from the four diagrams in Figure~\ref{fig:gravphoto} are given in Appendix~\ref{sec:gravscat} and have a rather complex form.  However, when added together we find a {\it much} simpler result.  The full graviton photo-production amplitude is found to be proportional to the already calculated Compton amplitude for spin-1---Eq.~\eqref{eq:gh}---times a universal kinematic factor.  That is,
\begin{equation}
\big\langle p_f;k_f,\epsilon_f\epsilon_f\big|T\big|p_i;k_i,\epsilon_i\big\rangle=\sum_{i=a,b,c,d}{\rm Amp}\,({\rm Fig}. 4(i))=H\times
\left(\epsilon_{f\alpha}^*\epsilon_{i\beta}
T_{\rm Compton}^{\alpha\beta}(S=1)\right)\,,\label{eq:gi}
\end{equation}
where
\begin{equation}
H={\kappa\over 2e}{p_f\cdot F_f\cdot p_i\over k_i\cdot k_f}=
{\kappa\over 2e}\,{\epsilon_f^*\cdot p_f\,k_f\cdot
p_i-\epsilon_f^*\cdot p_i\,k_f\cdot p_f\over k_i\cdot k_f}\,,
\end{equation}
and $\epsilon_{f\alpha}^*\epsilon_{i\beta} T_{\rm Compton}^{\alpha\beta}(S)$ is the spin-1 Compton scattering amplitude
calculated in the previous section.  The gravitational and electromagnetic
gauge invariance of Eq.~(\ref{eq:gi}) is obvious, since it follows directly from the gauge invariance already shown for the Compton amplitude together with the explicit gauge invariance of the factor $H$.  The validity of Eq.~(\ref{eq:gi}) allows the straightforward calculation of the cross-section by helicity methods since the graviton photo-production helicity amplitudes are given simply by
\begin{equation}
C^1(ab;cd)=H\times B^1(ab;cd)\,,
\end{equation}
where $B^1(ab;cd)$ are the Compton helicity amplitudes found in the previous section.  We can then evaluate the invariant photo-production
cross-section using
\begin{equation}
{d\sigma^{\rm photo}_{S=1}\over dt}={1\over 16\pi\big(s-m^2\big)^2}{1\over
3}\sum_{a=-,0,+}{1\over 2}\sum_{c=-,+}\big|C^1(ab;cd)\big|^2\,,\label{eq:ph}
\end{equation}
yielding
\begin{eqnarray}
{d\sigma^{\rm photo}_{S=1}\over dt}&=&-{e^2\kappa^2(m^4-su)\over 96\pi t\big(s-m^2\big)^4\big(u-m^2\big)^2}\left.\Big[(m^4-su+t^2)\big(3(m^4-su)+t^2\big)\right.\nonumber\\
&+&\left.t^2(t-m^2)(t-3m^2)\right.\Big]\,.
\end{eqnarray}

Since
\begin{equation}
|H|={\kappa\over e}\left({m^4-su\over -2t}\right)^{1\over 2}\,,
\end{equation}
the laboratory value of the factor $H$ is
\begin{equation}
|H_{lab}|^2={\kappa^2m^2\over 2e^2}{\cos^2{1\over 2}\theta_L\over
\sin^2{1\over 2}\theta_L}\,,
\end{equation}
and the corresponding laboratory cross-section is
\begin{eqnarray}
{d\sigma_{{\rm lab},S=1}^{\rm photo}\over
  d\Omega}&=&\big|H_{lab}\big|^2{d\sigma^{\rm Comp}_{{\rm lab},S=1}\over
  dt}\nonumber\\
&=&G\alpha\cos^2{\theta_L\over 2}\left({\omega_f\over \omega_i}\right)^4\left[\bigg({\rm ctn}^2{\theta_L\over 2}\cos^2{\theta_L\over 2}+\sin^2{\theta_L\over 2}\bigg)\bigg(1+2{\omega_i\over m}\sin^2{\theta_L\over 2}\bigg)^2\right.\nonumber\\
&+&\left.{16\omega_i^2\over 3m^2}\sin^2{\theta_L\over 2}\bigg(1+2{\omega_i\over m}\sin^2{\theta_L\over 2}\bigg)+{32\omega_i^4\over 3m^4}\sin^6{\theta_L\over 2}\right]\,.\label{eq:gb}
\end{eqnarray}

Comparing Eq. (\ref{eq:gb}) with the spin-0 and spin-${1\over 2}$ cross sections found in ~\cite{ppa}
\begin{eqnarray}
{d\sigma_{{\rm lab},S=0}^{\rm photo}\over  d\Omega}&=&G\alpha\cos^2{\theta_L\over 2}\left({\omega_f\over \omega_i}\right)^2\left[{\rm ctn}^2{\theta_L\over 2}\cos^2{\theta_L\over 2}+\sin^2{\theta_L\over 2}\right]\nonumber\\
{d\sigma_{{\rm lab},S={1\over 2}}^{\rm photo}\over  d\Omega}&=&G\alpha\cos^2{\theta_L\over 2}\left({\omega_f\over \omega_i}\right)^3\left[\left({\rm ctn}^2{\theta_L\over 2}\cos^2{\theta_L\over 2}+\sin^2{\theta_L\over 2}\right)
+{2\omega_i\over m}\left(\cos^4{\theta_L\over 2}+\sin^4{\theta_L\over 2}\right)\right.\nonumber\\
&+&\left.2{\omega_i^2\over m^2}\sin^2{\theta_L\over 2}\right],
\end{eqnarray}
we see that, just as in Compton scattering, the low-energy laboratory cross-section has a universal form, which is valid for a target of arbitrary spin
\begin{equation}
{d\sigma_{{\rm lab},S}^{\rm photo}\over d\Omega}=G\alpha\cos^2{\theta_L\over 2}\bigg({\rm ctn}^2{\theta_L\over 2}\cos^2{\theta_L\over 2}+\sin^2{\theta_L\over 2}\bigg)\bigg(1+{\cal O}\Big({\omega_i\over m}\Big)\bigg)\,.\label{eq:bz}
\end{equation}
In this case the universality can be understood from the feature that at low-energy the leading contribution to the graviton photo-production amplitude comes {\it not} from the seagull, as in Compton scattering, but rather from the photon pole term,
\begin{equation}
{\rm Amp}_{\gamma-{\rm pole}}\ \ \underset{\omega\ll m}{\longrightarrow}\ \ \kappa{\epsilon_f^*\cdot\epsilon_i\,\epsilon_f^*\cdot k_i\over 2k_f\cdot k_i}\times k_i^\mu\,\big\langle p_f;S,M_f\big|J_\mu\big|p_i;S,M_i\big\rangle\,.
\end{equation}
The leading piece of the electromagnetic current has the universal low-energy structure
\begin{equation}
\big\langle p_f;S,M_f\,\big|J_\mu\big|\,p_i;S,M_i\big\rangle={e\over 2m}\big(p_f+p_i\big)_\mu\delta_{M_f,M_i}\bigg(1+{\cal O}\Big({p_f-p_i\over m}\Big)\bigg)\,,
\end{equation}
where we have divided by the factor $2m$ to account for the normalization of the target particle.  Since $k_i\cdot(p_f+p_i)\underset{\omega\rightarrow 0}{\longrightarrow}2m\omega$, we find the universal low-energy amplitude
\begin{equation}\label{e:NRgpole}
{\rm Amp}_{\gamma-pole}^{NR}=\kappa \,e\, \omega\,{\epsilon_f^*\cdot\epsilon_i\,\epsilon_f^*\cdot k_i\over 2k_f\cdot k_i}\,,
\end{equation}
whereby the helicity amplitudes have the form
\begin{eqnarray}\displaystyle
{\rm Amp}_{\gamma-pole}^{NR}={\kappa \, e\over 2\sqrt{2}}\left\{\begin{array}{ll}
{1\over 2}\sin\theta_L\left.\Big({1+\cos\theta_L\over 1-\cos\theta_L}\right.\Big)={\cos{\theta_L\over 2}\over\sin{\theta_L\over 2}}\cos^2{\theta_L\over 2}& ++=--\,,\\[2ex]
{1\over 2}\sin\theta_L\left.\Big({1-\cos\theta_L\over 1-\cos\theta_L}\right.\Big)={\cos{\theta_L\over 2}\over \sin{\theta\over 2}}\sin^2{\theta_L\over 2}&+-=-+\,.
\end{array}\right.
\end{eqnarray}
Squaring and averaging, summing over initial, final spins we find then
\begin{equation}
{d\sigma_{{\rm lab},S}^{\rm photo}\over d\Omega}\ \
\underset{\omega\ll m}{\longrightarrow}\ \ G\,\alpha\cos^2{\theta_L\over 2}\left[\bigg({\rm ctn}^2{\theta_L\over 2}\cos^2{\theta_L\over 2}+\sin^2{\theta_L\over 2}\bigg)\left(1+{\cal O}\left({\omega_i\over m}\right)\right)\right]\,,
\end{equation}
as determined above---{\it cf.} Eq. (\ref{eq:bz}).

Comparing the individual contributions from the Appendix~\ref{sec:gravscat} with the simple forms above, the power of factorization is obvious and, as we shall see in the next section, permits the straightforward evaluation of even more complex reactions such as gravitational Compton scattering.\\

\subsection{Gravitational Compton Scattering}\label{sec:gravcomp}

\begin{figure}[h]
\begin{center}
\includegraphics[width=10cm]{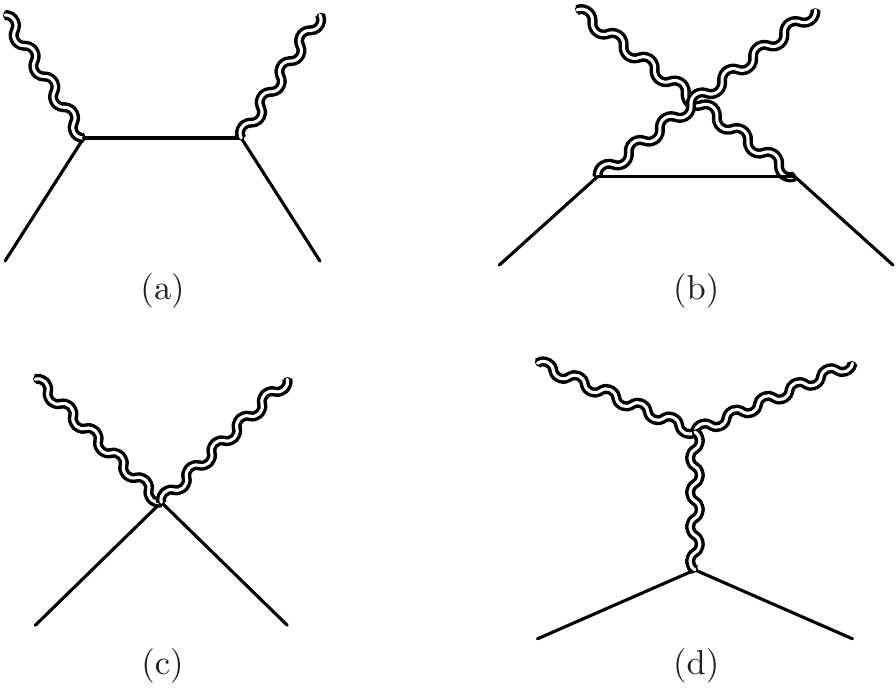}
\caption{Diagrams
relevant for gravitational Compton scattering. }\label{fig:gravcomp}
\end{center}
\end{figure}

In the previous section we observed the power of factorization in the context of graviton photo-production on a spin-1 target
in that we only needed to calculate the simpler Compton scattering
process rather than to consider the full gravitational interaction description.
In this section we consider an even more challenging example, that of
gravitational Compton scattering---$g+S\rightarrow g+S$---from a
spin-1 target, for which there exist the four diagrams shown in
Figure~\ref{fig:gravcomp}.

The contributions from the four individual diagrams can be calculated using the
graviton vertices given above and are
quoted in Appendix~\ref{sec:gravscat}.  Each of the four diagrams has a rather complex
form.  However, when added together the total again simplifies
enormously. Defining the kinematic factor
\begin{equation}
Y={\kappa^2\over 8e^4}{p_i\cdot k_i\,p_i\cdot k_f\over k_i\cdot k_f}={\kappa^4\over 16e^4}{(s-m^2)\,(u-m^2)\over t}\,,
\end{equation}
the sum of the four diagrams is found to be
\begin{eqnarray}
&&\hskip-2.4cm\big\langle p_f,\epsilon_B;k_f,\epsilon_f\epsilon_f\,\big|{\rm Amp}_{\rm grav}\big|\,p_i,\epsilon_A;k_i,\epsilon_i\epsilon_i\big\rangle_{S}=\sum_{i=1}^4{\rm Amp}\,({\rm Fig.}5(i))\nonumber\\
&=&Y\times]\big\langle p_f,\epsilon_B;k_i,\epsilon_f\,\big|
{\rm Amp}_{\rm em}\big|\,p_i,\epsilon_A;k_i,\epsilon_i\big\rangle_{S}
\times\big\langle p_f;k_i,\epsilon_f\,\big|{\rm Amp}_{\rm em}\big|\,p_i;k_i,\epsilon_i\rangle_{S=0}\,,\nonumber\\
\quad\label{eq:cd}
\end{eqnarray}
with $S=1$, where
\begin{equation}
\langle p_f;k_i,\epsilon_f|{\rm Amp}_{\rm em}|p_i;k_i,\epsilon_i\rangle_{S=0}=2e^2\left[{\epsilon_i\cdot p_i\,\epsilon_f^*\cdot p_f\over p_i\cdot k_i}-{\epsilon_i\cdot p_f\,\epsilon_f^*\cdot p_i\over p_i\cdot k_f}-\epsilon_f^*\cdot\epsilon_i\right]\,,\label{eq:vg}
\end{equation}
is the Compton amplitude for a spinless target.

In ~\cite{ppa} the identity Eq.~\eqref{eq:cd} was verified for simpler cases of $S=0$ and $S={1\over 2}$. This relation is a
consequence of the general connections between gravity and gauge theory tree-level amplitudes derived using string-based methods  as explained in~\cite{bxa}.  Here we have demonstrated its validity for the much more complex case of spin-1 scattering.
The corresponding cross-section can be calculated by helicity methods using
\begin{equation}
D^1(ab;cd)=Y\times B^1(ab;cd)\times A^0(cd)\,,
\end{equation}
where $D^1(ab;cd)$ is the spin-1 helicity amplitude for gravitational Compton scattering, $B^1(ab;cd)$ is the ordinary spin-1 Compton helicity amplitude calculated in section~\ref{sec:compton}, and
\begin{eqnarray}
A^0(++)&=&2e^2{m^4-su\over \big(s-m^2)\big(u-m^2)}\,,\nonumber\\
A^0(+-)&=&2e^2{-m^2t\over \big(s-m^2)\big(u-m^2)}\,,
\end{eqnarray}
are the helicity amplitudes for spin zero Compton scattering rising from Eq. \ref{eq:vg} calculated in ~\cite{ppa}.
Using Eq. (\ref{eq:cd}) the invariant cross-section for unpolarized spin-1 gravitational Compton scattering
\begin{equation}
{d\sigma^{\rm g-Comp}_{S=1}\over dt}={1\over 16\pi\big(s-m^2\big)^2}\ { 1\over
3}\sum_{a=-,0,+}{1\over 2}\sum_{c=-,+}\big|D^1(ab;cd)\big|^2\,,\label{eq:mh}
\end{equation}
is found to be
\begin{eqnarray}
{d\sigma^{\rm g-Comp}_{S=1}\over dt}&=&{\kappa^4\over 768\pi\big(s-m^2\big)^4\big(u-m^2\big)^2t^2}
\left[(m^4-su)^2\big(3(m^4-su)+t^2)(m^4-su+t^2)\big)\right.\nonumber\\
&+&\left.m^4t^4(3m^2-t)(m^2-t)\right]\,,
\end{eqnarray}
and this form can be compared with the unpolarized gravitational Compton cross-sections found in ~\cite{ppa}
\begin{eqnarray}
{d\sigma^{\rm g-Comp}_{S=0}\over d\Omega}&=&{\kappa^4\over 256\pi\big(s-m^2\big)^4\big(u-m^2\big)^2t^2}\left.\Big[(m^4-su)^4+m^8t^4\right.\Big]\,\nonumber\\
{d\sigma^{\rm g-Comp}_{S={1\over 2}}\over dt}&=&{\kappa^4\over 512\pi\big(s-m^2\big)^4\big(u-m^2t^2\big)^2}\left[\big(m^4-su\big)^3\big(2(m^4-su)+t^2\big)+m^6t^4\big(2m^2-t\big)\right]\,.\nonumber\\
\quad
\end{eqnarray}
The corresponding laboratory frame cross-sections are
\begin{eqnarray}
{d\sigma_{{\rm lab},S=1}^{\rm g-Comp}\over d\Omega}&=&G^2m^2{\omega_f^4\over \omega_i^4}\left[\bigg({\rm ctn}^4{\theta_L\over 2}\cos^4{\theta_L\over 2}+\sin^4{\theta_L\over 2}\bigg)\bigg(1+2{\omega_i\over m}\sin^2{\theta_L\over 2}\bigg)^2\right.\nonumber\\
&+&\left.{16\over 3}{\omega_i^2\over m^2}\bigg(\cos^6{\theta_L\over 2}+\sin^6{\theta_L\over 2}\bigg)\bigg(1+2{\omega_i\over m}\sin^2{\theta_L\over 2}\bigg)\right.\nonumber\\
&+&\left.{16\over 3}{\omega_i^4\over m^4}\sin^2{\theta_L\over 2}\bigg(\cos^4{\theta_L\over 2}+\sin^4{\theta_L\over 2}\bigg)\right]\,,\nonumber\\
{d\sigma^{\rm g-Comp}_{{\rm lab},S={1\over 2}}\over d\Omega}&=&G^2m^2{\omega_f^3\over \omega_i^3}\left[\bigg({\rm ctn}^4{\theta_L\over 2}\cos^4{\theta_L\over 2}+\sin^4{\theta_L\over 2}\bigg)+2{\omega_i\over m}\bigg({\rm ctn}^2{\theta_L\over 2}\cos^6{\theta_L\over 2}+\sin^6{\theta_L\over 2}\bigg)\right.\nonumber\\
&+&\left.2{\omega_i^2\over m^2}\bigg(\cos^6{\theta_L\over 2}+\sin^{6}{\theta_L\over 2}\bigg)\right]\,,\nonumber\\
{d\sigma^{\rm g-Comp}_{{\rm lab},S=0}\over d\Omega}&=&G^2m^2{\omega_f^2\over\omega_i^2}\left[{\rm ctn}^4{\theta_L\over 2}\cos^4{\theta_L\over 2}+\sin^4{\theta_L\over 2}\right]\,.\emph{}
\end{eqnarray}
We observe that the low-energy laboratory cross-section has the universal form for any spin
\begin{equation}
{d\sigma_{{\rm lab},S}^{\rm g-Comp}\over d\Omega}=G^2m^2\left[\left({\rm ctn}^4{\theta_L\over 2}\cos^4{\theta_L\over 2}+\sin^4{\theta_L\over 2}\right)\left(1+{\cal O}\Big({\omega_i\over m}\Big)\right)\right]\,.\label{eq:zs}
\end{equation}

It is interesting to note that the ``dressing" factor for the leading $(++)$ helicity Compton amplitude---
\begin{equation}
\big |Y\big|\,\big|A^{++}\big|\ =\ {\kappa^2\over 2e^2}\ {m^4-su\over -t}\ \ \stackrel{\rm lab}{\longrightarrow}\ \ {\kappa^2m^2\over 2e^2}\ {\cos^2{\theta_L\over 2}\over \sin^2{\theta_L\over 2}}\,,
\end{equation}
---is simply the square of the photo-production dressing factor $H$, as might intuitively be expected since now {\it both} photons must be dressed in going from the Compton to the gravitational Compton cross-section.\footnote{In the case of $(+-)$ helicity the ``dressing" factor is
\begin{equation}
\big|Y\big|\,\big|A^{+-}\big|={\kappa^2\over 2e^2}\ m^2\,,
\end{equation}
so that the non-leading contributions will have different dressing factors.}
In this case the universality of the nonrelativistic cross-section follows from the leading contribution arising from the graviton pole term
\begin{equation}
{\rm Amp}_{g-pole}\ \ \underset{\omega\ll m}{\longrightarrow}\ \
{\kappa\over 4k_f\cdot k_i}\ \big(\epsilon_f^*\cdot\epsilon_i\big)^2\ \big(k_f^\mu k_f^\nu+k_i^\mu k_i^\nu\big)\ {\kappa\over 2}\ \big \langle p_f;S,M_f\,\big|T_{\mu\nu}\big |\,p_i;S,M_i\big\rangle\,.
\end{equation}
Here the matrix element of the energy-momentum tensor has the universal low-energy structure
\begin{equation}
{\kappa\over 2}\,\big \langle p_f;S,M_f\,\big|T_{\mu\nu}\big |\,p_i;S,M_i\big\rangle={\kappa\over 4m}\,\big(p_{f\mu}p_{i\nu}+p_{f\nu}p_{i\mu}\big)\,\delta_{M_f,M_i}\Big(1+{\cal O}\Big({p_f-p_i\over m}\Big)\Big)\,,
\end{equation}
where we have divided by the factor $2m$ to account for the normalization of the target particle.  We find then the universal form for the leading graviton pole amplitude
\begin{equation}
{\rm Amp}_{g-pole}\ \ \underset{\rm non-rel}{\longrightarrow}\ \
{\kappa^2\over 8m\, k_f\cdot k_i}\ \big(\epsilon_f^*\cdot\epsilon_i\big)^2\,\big(p_i\cdot k_f\,p_f\cdot k_f+p_i\cdot k_i\, p_f\cdot k_i\big)\,\delta_{M_f,M_i}\,.
\end{equation}
Since $p\cdot k\underset{\omega\ll m}{\longrightarrow}m\omega$ the helicity amplitudes become
\begin{eqnarray}
{\rm Amp}_{g-pole}^{\rm NR}=4\pi Gm\left\{\begin{array}{ll}
{\big(1+\cos\theta_L\big)^2\over 2\big(1-\cos\theta_L\big)}={\cos^4{\theta_L\over 2}\over \sin^2{\theta_L\over 2}}& ++=--\,,\\[2ex]
{\big(1-\cos\theta_L\big)^2\over 2\big(1-\cos\theta_L\big)}={\sin^4{\theta_L\over 2}\over \sin^2{\theta_L\over 2}}&+-=-+\,.
\end{array}\right.
\end{eqnarray}
Squaring and averaging,summing over initial,final spins we find
\begin{equation}
{d\sigma_{{\rm lab},S}^{\rm g-Comp}\over d\Omega}\ \
\underset{\omega\ll m}{\longrightarrow}\ \ G^2\,m^2\,\bigg[{\rm ctn}^4{\theta_L\over 2}\cos^4{\theta_L\over 2}+\sin^4{\theta_L\over 2}\bigg]\,,
\end{equation}
as found in Eq.~\eqref{eq:zs} above.

\section{Graviton-Photon Scattering}\label{sec:gravpho}
In the previous sections we have generalized the results of ~\cite{ppa} to the case of a massive spin-1 target.  Here we show how these spin-1 results can be used to calculate the cross-section for photon-graviton scattering.  In the Compton scattering calculation we assumed that the spin-1 target had charge $e$.  However, the photon couplings to the graviton are identical to those of a graviton coupled to a charged spin-1 system in the massless limit, and one might assume then that, since the results of the gravitational Compton scattering are independent of charge, the graviton-photon cross-section can be calculated by simply taking the $m\rightarrow 0$ limit of the graviton-spin-1 cross-section.  Of course, the laboratory cross-section no longer makes sense since the photon cannot be brought to rest, but the invariant cross-section is well defined
\begin{equation}
{d\sigma_{S=1}^{\rm g-Comp}\over dt}\ \ \underset{m\rightarrow 0}{\longrightarrow}\ \ {4\pi \,G^2\,\big(3s^2u^2-4t^2su+t^4\big)\over 3s^2t^2}={4\pi G^2(s^4+u^4+s^2u^2)\over 3s^2t^2}\,,\label{eq:hd}
\end{equation}
and it might be (naively) assumed that Eq.~(\ref{eq:hd}) is the graviton-photon scattering cross-section.  However, this is {\it not} the case and the resolution of this problem involves some interesting physics.

We begin by noting that in the massless limit the only nonvanishing helicity amplitudes are
\begin{eqnarray}
D^1(++;++)_{m=0}&=&D^1(--;--)_{m=0}=8\pi\, G\,{s^2\over t}\,,\nonumber\\
D^1(--;++)_{m=0}&=&D^1(++;--)_{m=0}=8\pi\, G\,{u^2\over t}\,,\nonumber\\
D^1(00;++)_{m=0}&=&D^1(00;--)_{m=0} \ \,=8\pi\, G\,{su\over t}\,,
\end{eqnarray}
which lead to the cross-section
\begin{eqnarray}
{d\sigma^{\rm g-Comp}_{S=1}\over dt}&=& {1\over 16\pi s^2}\,\frac13\,\sum_{a=+,0,-}\frac12\sum_{c=+,-}\big|D^1(ab;cd)\big|^2\cr
&=&{1\over 16\pi s^2}\,{1\over 3\cdot 2}\,(8\pi G)^2\times
2\times\left[{s^4\over t^2}+{u^4\over t^2}+{s^2u^2\over t^2}\right]\cr
&=&{4\pi\over 3}\,G^2{s^4+u^4+s^2u^2\over s^2t^2}\,,
\end{eqnarray}
in agreement with Eq. (\ref{eq:hd}).  However, this result reveals the problem.  We know that in Coulomb gauge the photon has only {\it two} transverse degrees of freedom, corresponding to positive and negative helicity---there exists {\it no} longitudinal degree of freedom.  Thus the correct photon-graviton cross-section is obtained by deleting the contribution from the $D^1(00;++)$ and $D^1(00;--)$ multipoles
\begin{eqnarray}
{d\sigma_{g\gamma}\over dt}&=& {1\over 16\pi s^2}\,\frac13\sum_{a=+,-}\frac12\sum_{c=+,-}\big|D^1(ab;cd)\big|^2\cr
&=&{1\over 16\pi s^2}\,{1\over 2\cdot 2}\,\big(8\pi G\big)^2\times 2\times\left[{s^4\over t^2}+{u^4\over t^2}\right]=2\pi\, G^2\,{s^4+u^4\over s^2t^2}\,,
\end{eqnarray}
which agrees with the value calculated via conventional methods by Skobelev~\cite{skb}.  Alternatively, since in the center of mass frame
\begin{equation}
{dt\over d\Omega}={\omega_{\rm CM}\over \pi}\,,
\end{equation}
we can write the center of mass graviton-photon cross-section in the form
\begin{equation}\label{e:gG}
{d\sigma_{\rm CM}\over d\Omega}=2\,G^2\,\omega_{\rm CM}^2\left({1+\cos^8{\theta_{\rm CM}\over 2}\over \sin^4{\theta_{\rm CM}\over 2}}\right)\,,
\end{equation}
again in agreement with the value given by Skobelev~\cite{skb}.

So what has gone wrong here?  Ordinarily in the massless limit of a spin-1 system, the longitudinal mode decouples because the zero helicity spin-1 polarization vector becomes
\begin{equation}
\epsilon_\mu^0\ \ \underset{m\rightarrow 0}{\longrightarrow}\ \ {1\over m}\bigg(p,\Big(p+{m^2\over 2p}+\ldots\Big)\hat{z}\bigg)={1\over m}p_\mu+\Big(0,{m\over 2p}\hat{z}\Big)+\ldots\label{eq:fc}
\end{equation}
However, the term proportional to $p_\mu$ vanishes when contracted with a conserved current by gauge invariance while the term in ${m\over 2p}$ vanishes in the massless limit.  Indeed that the spin-1 Compton scattering amplitude becomes gauge invariant for a massless spin-1 system can be seen from the fact that the Compton amplitude can be written as
\begin{eqnarray}
&&\hskip-1.4cm{\rm Amp}^{\rm Comp}_{S=1}\underset{m\rightarrow 0}{\longrightarrow}{e^2\over p_i\cdot q_i\,p_i\cdot q_f}\left.\bigg[{\rm Tr}\big(F_iF_fF_AF_B\big)+{\rm Tr}\big(F_iF_AF_fF_B\big)+{\rm Tr}\big(F_iF_AF_BF_f\big)\right.\nonumber\\
&&\hskip1.5cm-\left.{1\over 4}\left.\Big({\rm Tr}\big(F_iF_f\big){\rm Tr}\big(F_AF_B\big)
+{\rm Tr}\big(F_iF_A\big){\rm Tr}\big(F_fF_B\big)+{\rm Tr}\big(F_iF_B\big){\rm Tr}\big(F_fF_A\big)\right.\Big)\right]\,,\nonumber\\
\quad
\end{eqnarray}
which can be checked by a bit of algebra.  Equivalently, one can verify explicitly that the massless spin-1 amplitude vanishes if one replaces either $\epsilon_{A\mu}$ by $p_{i\mu}$ {\it or} $\epsilon_{B\mu}$ by $p_{f\mu}$.  However, what takes place when {\it two} longitudinal spin-1 particles are present is that the product of longitudinal polarization vectors is proportional to $1/m^2$, while the correction term to the four-momentum $p_\mu$ is ${\cal O}(m^2)$ so that the product is nonvanishing in the massless limit and this is why the multipole $D(00;++)_{m=0}=D(00;--)_{m=0}$ is nonzero.  One deals with this problem by simply omitting the longitudinal degree of freedom explicitly, as done above.

\subsection{Extra Credit}

Before leaving this section it is interesting to note that graviton-graviton scattering can be treated in a parallel fashion.  That is, the graviton-graviton scattering amplitude can be obtained by dressing the product of two massless spin-1 Compton amplitudes~\cite{str}---
\begin{align}
&\big\langle p_f,\epsilon_B\epsilon_B;k_f,\epsilon_f\epsilon_f\,\big|{\rm Amp}^{\rm tot}_{grav}\big|\,p_i\epsilon_A\epsilon_A;k_i,\epsilon_i\epsilon_i\big\rangle_{m=0,S=2}=\\
&Y\times\langle p_f,\epsilon_B;k_f,\epsilon_f\,|{\rm Amp}^{\rm Comp}_{em}|\,p_i,\epsilon_A;k_i\epsilon_i\rangle_{m=0,S=1}
\times\langle p_f,\epsilon_B;k_f,\epsilon_f\,|{\rm Amp}^{\rm Comp}_{em}|\,p_i,\epsilon_A;k_i\epsilon_i\rangle_{m=0,S=1}\,.\nonumber
\end{align}
Then for the helicity amplitudes we have
\begin{equation}
E^2(++;++)_{m=0}=Y\big(B^1(++;++)_{m=0}\big)^2\,,
\end{equation}
where $E^2(++;++)$ is the graviton-graviton $++;++$ helicity amplitude while $B^1(++;++)$ is the corresponding spin-1 Compton helicity amplitude.  Thus we find
\begin{equation}
E^2(++;++)_{m=0}={\kappa^2\over 16e^4}\,{su\over t}\times \bigg({2e^2\,{s\over u}}\bigg)^2=8\pi \,G\,{s^3\over ut}\,,
\end{equation}
which agrees with the result calculated via conventional methods~\cite{pfg}.

\section{The forward cross-section}\label{sec:forward}

It is interesting to note some intriguing physics associated with the forward-scattering limit.  In this limit, {\it i.e.}, $\theta_L\to 0$, in the laboratory frame, the Compton cross-section evaluated in section~\ref{sec:compton}
has a universal structure independent of the spin $S$ of the massive target
\begin{equation}\label{e:sigmaCompton}
\lim_{\theta_L\to 0}   {d\sigma^{\rm Comp}_{{\rm lab},S}\over d\Omega} ={\alpha^2\over2m^2}\,,
\end{equation}
reproducing the well-known Thomson scattering cross-section.

\medskip
For graviton photo-production, however, the small-angle limit is very different,
since the forward-scattering cross-section is divergent---the small angle limit of the graviton photo-production of
section~\ref{sec:gravphotoprod} is given by

\begin{equation} \label{e:fg}
  \lim_{\theta_L\to0}   {d\sigma^{\rm photo}_{{\rm lab},S}\over d\Omega} = {4G\alpha\over \theta_L^2}\,,
\end{equation}
and arises from the photon pole in figure~\ref{fig:gravphoto}(d). Notice that this behavior {\it differs} from
the familiar $1/\theta^4$ small-angle Rutherford cross-section for scattering in a Coulomb-like potential.  Rather, this divergence of the forward cross-section indicates that a long range force is involved but with an effective $1/r^2$ potential. This effective potential arising from the $\gamma$-pole in figure~\ref{fig:gravphoto}(d), is the Fourier transform with respect to the momentum transfer $q=k_f-k_i$ of the low-energy limit given in Eq.~\ref{e:NRgpole}. Because of the linear dependence in the momenta in the numerator, one obtains
\begin{equation}
\int {d^3 q\over (2\pi)^3}\,  e^{i \boldsymbol{q}\cdot \boldsymbol{r}} \,
  {1\over |\boldsymbol{q}|} = {1\over 2\pi^2 r^2}\,,
\end{equation}
and this result is the origin of the peculiar forward-scattering behavior of the cross-section.
Another contrasting feature of graviton photo-production is the independence of the forward cross-section on the mass $m$ of the target.

\medskip

The small angle limit of the gravitational Compton cross-section derived in section~\ref{sec:gravcomp} is given by

\begin{equation}\label{e:fG}
  \lim_{\theta_L\to0}
  {d\sigma^{\rm g-Comp}_{lab,S}\over d\Omega} = {16G^2m^2\over \theta_L^4}\,,
\end{equation}
where again the limit is independent of the spin $S$ of the matter field.  Finally, the photon-graviton cross-section derived in
section~\ref{sec:gravpho}, has the forward-scattering dependence
\begin{equation}\label{e:fG2}
  \lim_{\theta_{\rm CM}\to0} {d\sigma_{\rm CM}\over d\Omega}=
  {32G^2\omega_{\rm CM}^2\over \theta_{\rm CM}^4}  \,.
\end{equation}
The behaviors in Eqs. (\ref{e:fG}) and (\ref{e:fG2}) are due to the graviton pole in figure~\ref{fig:gravcomp}(d), and are typical of the small-angle behavior of Rutherford scattering in a Coulomb potential.

The classical bending of the geodesic for a massless particle in a Schwarzschild metric produced by a point-like mass $m$ is given by $b=4Gm/\theta+O(1)$~\cite{dva}, where $b$ is the classical impact parameter.  The associated classical cross-section is
\begin{equation}
{  d\sigma^{\rm classical}  \over d\Omega}= {b\over \sin\theta}
\left|db\over d\theta\right| \simeq {16G^2m^2\over \theta^4}+O(\theta^{-3})\,,
\end{equation}
matching the expression in Eq. (\ref{e:fG}).  The diagram in Figure \ref{fig:gravcomp}(d) describes the gravitational interaction between a massive particle of spin-$S$ and a graviton. In the forward-scattering limit the remaining diagrams of figure \ref{fig:gravcomp} have vanishing contributions.  Since this limit is independent of the spin of the particles interacting
gravitationally, the expression in Eq. (\ref{e:fG}) describes the forward gravitational scattering cross-section of \emph{any} massless particle on the target of mass $m$ and explains the match with the classical formula given above.

Eq. (\ref{e:fG2}) can be interpreted in a similar way, as the bending of a geodesic in a geometry curved by the energy density with an effective Schwarzschild radius of $\sqrt2\,G\omega_{\rm CM}$ determined by the center-of-mass energy~\cite{ama}.
However, the effect is fantastically small since the cross-section in Eq. (\ref{e:fG2}) is of order $\ell_P^4/(\lambda^2\, \theta_{\rm CM}^4)$ where $\ell_P^2=\hbar G/c^3\sim1.62\,10^{-35}\,$m is the Planck length, and $\lambda$ the wavelength of the photon.

\section{Bending of Light in Classical General Relativity}

We close our discussion by examining the process of gravitational light bending and look at different pictures by which this phenomenon can be discussed.  We begin with the standard general relaticistic derivation.  The theory of general relativity encapsulates the theory of gravity in terms of a simple second rank tensor equation~\cite{Wei72}
\begin{equation}
R_{\mu\nu}-{1\over 2}g_{\mu\nu}R=-{\kappa^2\over 4}T_{\mu\nu}\,,
\end{equation}
where $\kappa^2=32\pi G$ is the gravitational coupling constant, $T_{\mu\nu}$ is the energy-momentum tensor,
\begin{equation}g_{\mu\nu}=\eta_{\mu\nu}+h_{\mu\nu}\,,\end{equation} is the metric tensor, and $R_{\mu\nu},\,R$ are the Ricci curvature tensor, scalar curvature, which are
defined in terms of $h_{\mu\nu}$ as
\begin{eqnarray}
R_{\mu\nu}&=&{\kappa\over 2}\left[\partial_\mu\partial_\nu
h+
\partial_\lambda\partial^\lambda
 h_{\mu\nu}-\partial_\mu\partial_\lambda
{h}^\lambda{}_\nu
-\partial_\nu\partial_\lambda {h}^\lambda{}_\mu\right]\,,\nonumber\\
R&=&\eta^{\mu\nu}R_{\mu\nu}=\kappa\left[\Box h-\partial_\mu\partial_\nu h^{\mu\nu}\right]\,,
\end{eqnarray}
in the linear approximation.  For a spatially isotropic spacetime the invariant time interval $d\tau$ is defined via a metric tensor of the form
$$d\tau^2=g_{\mu\nu}dx^\mu dx^\nu=A(r)dt^2-B(r)dr^2-r^2d\Omega^2\,,$$
and vanishes in the case of the motion of a massless system such as a photon.  We represent the sun as a simple non-spinning massive
object, described by the Schwarzschild metric~\cite{Sch16}
\begin{equation}
A(r)=1-{2GM\over r},\quad B(r)={1\over 1-{2GM\over r}}\,.
\end{equation}
The solution for the bending angle is then given by standard methods~\cite{Wei72}
\begin{equation}
\theta=2\int_0^1{du\over \sqrt{1-u^2-{2GM\over D}(1-u^3)}}-\pi\,,\label{eq:bv}
\end{equation}
where we have defined $u={D\over r}$.  Here $D$ is the distance of closest approach in Scwarzschild coordinates.  The integration in Eq. (\ref{eq:bv}) can be performed exactly in terms of elliptic functions, but
since near the solar rim $2GM/D\simeq 10^{-3}\ll 1$, we can instead use a perturbative solution
\begin{eqnarray}
\theta&=&2\int_0^1du\left[{1\over \sqrt{(1-u)(1+u)}}+{GM\over D}{1+u+u^2\over \sqrt{(1-u)(1+u)^3}}\right.\nonumber\\
&+&\left.{3\over 2}{G^2M^2\over D^2}{(1+u+u^2)^2\over \sqrt{(1-u)(1+u)^5}}+\ldots\right]-\pi\nonumber\\
&=&{4GM\over D}+{4G^2M^2\over D^2}\left({15\pi\over 16}-1\right)+\ldots
\end{eqnarray}
However, instead of using the coordinate-dependent quantity $D$, the bending angle should be written in terms of the invariant impact parameter $b$, defined as
\begin{equation}
b=\sqrt{B(D)}D={R\over \sqrt{1-{2GM\over D}}}=D+GM+\ldots\,,
\end{equation}
we have then
\begin{equation}
\theta={4GM\over b}+{15\pi\over 4}{G^2M^2\over b^2}+\ldots\,,
\end{equation}
which is the standard result, together with the next to
leading order correction.

\section{Quantum Mechanical Scattering Amplitude}

Both alternative methods require the quantum mechanical scattering amplitude for the gravitational interaction of neutral massive and massless systems.  For simplicity we take both to be spinless and therefore described by the simple Klein-Gordon Lagrangian~\cite{Bjo64}
\begin{equation}
{\cal L}={1\over 2}\left(g^{\mu\nu}\partial_\mu\phi\partial_\nu\phi-m^2\phi^2\right)\,,
\end{equation}
so that the energy-momentum tensor is given by
\begin{equation}
T_{\mu\nu}=\left(2{\delta\over \delta g^{\mu\nu}}-g_{\mu\nu}\right)\sqrt{-{\rm det}g}{\cal L}=\partial_\mu\phi\partial_\nu\phi-{1\over 2}\eta_{\mu\nu}\left(\partial_\alpha\phi\partial^\alpha\phi-m^2\phi^2\right)+\ldots
\end{equation}
The corresponding matrix element is then
\begin{equation}
<p_f|T_{\mu\nu}|p_i>=p_{f\mu} p_{i\nu}+p_{i\mu} p_{f\nu}-\eta_{\mu\nu}(p_f\cdot p_i-m^2)+\ldots
\end{equation}
Using the gravitational interaction~\cite{Sca72}
\begin{equation}
{\cal L}_{int}={\kappa\over 2}h^{\mu\nu}T_{\mu\nu}\,,
\end{equation}
and the harmonic gauge graviton propagator
\begin{equation}
D_F^{\alpha\beta;\gamma\delta}(q)={iP^{\alpha\beta;\gamma\delta}\over q^2}\,,
\end{equation}
where $P^{\alpha\beta;\gamma\delta}={1\over 2}(\eta^{\alpha\gamma}\eta^{\beta\delta}+\eta^{\alpha\delta}\eta^{\beta\gamma}-\eta^{\alpha\beta}\eta^{\gamma\delta})$,
the lowest order graviton exchange amplitude for interaction between massive and massless systems described via initial, final energy-momentum $P_i,\,P_f$ and
$p_i,\,p_f$ respectively, becomes ({\it cf.} Figure 1)
\begin{eqnarray}
i{\cal M}_0(q)&=&{1\over \sqrt{4E_fE_i}}{1\over \sqrt{4\epsilon_f\epsilon_i}}{\kappa\over 2}<P_f|T_{\alpha\beta}|P_i>D_F^{\alpha\beta;\gamma\delta}(q){\kappa\over 2}<p_f|T_{\gamma\delta}|p_i>\nonumber\\
&=&{\kappa^2\over 4}\left({s^2-2s(m^2+M^2)+m^4+M^4\over q^2}+s-M^2-m^2\right)\,,
\end{eqnarray}
where $s=(P_i+p_i)^2=(P_f+p_f)^2$, $q^2=(P_f-P_i)^2=(p_f-p_i)^2$, and we have divided by the conventional normalizing factors $\sqrt{2E}$ for each external scalar field.  Working in the rest frame of the system having mass $M$, if the light system has mass $m$, then in the nonrelativistic limit $s\simeq (M+m)^2$ and $q_0\simeq 0$ so that
\begin{equation}
{\cal M}_0(\boldsymbol{q})\simeq -{1\over 4Mm}{\kappa^2\over 4}{2M^2m^2\over \boldsymbol{q}^2}=-{4\pi GMm\over \boldsymbol{q}^2}\,.
\end{equation}

In Born approximation the transition amplitude is related to the potential via~\cite{Sak11}
\begin{equation}
{\cal M}_0(\boldsymbol{q})=<\boldsymbol{p}_f|\hat{V}_0|\boldsymbol{p}_i>=\int d^3r e^{i\boldsymbol{q}\cdot\boldsymbol{r}}V_0(\boldsymbol{r})\,,
\end{equation}
where $\boldsymbol{q}=\boldsymbol{p}_i-\boldsymbol{p}_f$.  The corresponding gravitational potential is then given by the inverse Fourier transform
\begin{equation}
V_0(r)=\int {d^3q\over (2\pi)^3}e^{-i\boldsymbol{q}\cdot\boldsymbol{r}}{\cal M}_0(\boldsymbol{q})=-{GMm\over r}\,,
\end{equation}
and has the expected Newtonian form.  However, if the light system is massless and carries energy $E_m$, then $s= M^2+2ME_m$ so that
\begin{equation}
{\cal M}_0(q)\simeq -{8\pi GME_m\over \boldsymbol{q}^2}\,.
\end{equation}
and the gravitational potential becomes\footnote{Here the factor of two difference between the massless and massive systems under the replacement $m\rightarrow E_m$ represents the well-known relation between the predicted light bending in the Newtonian and Einstein formulations of gravity~\cite{Oha13}.}
\begin{equation}
V_0(r)=\int {d^3q\over (2\pi)^3}e^{-i\boldsymbol{q}\cdot\boldsymbol{r}}{\cal M}_0(\boldsymbol{q})=-{2GME_m\over r}
\end{equation}
Higher order corrections to this lowest order potential can be found
by calculating loop effects.  Such calculations have been performed by
a number of groups and the next to
leading order (${\cal O}(G^2)$) form of the gravitational interaction amplitude has been found to be~\cite{Bje03}
\begin{equation}
{\cal M}_1(q)=-{15\pi^2 G^2M^2E_m\over 2\sqrt{\boldsymbol{q}^2}}+\ldots\,,
\end{equation}
which will be used below.  With these forms in hand, we can now proceed to alternative light bending calculations.

\section{Geometrical Optics}

\noindent In section 7 we presented the conventional (particle)
derivation of the bending angle, in terms of the trajectory traveled
by photons.  In this section we review an alternative approach,
presented in~\cite{Bjerrum-Bohr:2016hpa},  to describe the propagation of light via geometrical optics, wherein the beam travels through a region defined by a position-dependent index of refraction $n(r)$, in which case we have the equation of motion~\cite{Bra04}
\begin{equation}
{d\over ds}n{d\boldsymbol{r}\over ds}=\boldsymbol{\nabla}n\,,\label{eq:vf}
\end{equation}
where $\boldsymbol{r}(s)$ is the trajectory as a function of the path length $s$.  In the case of light $ds\simeq cdt$
and Eq. (\ref{eq:vf}) simplifies to
\begin{equation}
{1\over c^2}{d^2\boldsymbol{r}\over dt^2}={1\over n}\boldsymbol{\nabla}n\,.\label{eq:jn}
\end{equation}
The index of refraction describes the propagation of light when $E\neq |\boldsymbol{p}|$, and is defined by $n=E/|\boldsymbol{p}|$.  In our case, as discussed in the previous section, the presence of the gravitational interaction between photons and the sun leads to a modification of the energy and therefore to an effective position dependent index of refraction
\begin{equation}
n(r)\simeq (E_m-V(r))/E_m=1-{1\over E_m}V(r)\,.
\end{equation}
where, for a massless scalar with energy $E_m$ interacting with a mass $M$ we found\footnote{This description in terms of a position-dependent index of refraction has been called the optical-mechanical\break analogy and is given in terms of~\cite{Als98}
\begin{equation}
n(r)=\sqrt{B(r)/A(r)}=1+{2GM\over r}+\ldots
\end{equation}}
\begin{equation}
V_0(r)=-{2GME_m\over r}
\end{equation}
In the absence of a potential ($n(r)=1$) consider a light beam incident along the $\hat{\boldsymbol{e}}_y$ direction and with impact parameter $b$ on a massive target located at the origin.  The trajectory is then characterized by the straight line
\begin{equation}
\boldsymbol{r}_0(t)=b\hat{\boldsymbol{e}}_x+ct\hat{\boldsymbol{e}}_y,\qquad -\infty<t<\infty .
\end{equation}
If we now turn on a potential $V_0(r)$ the index of refraction is no longer unity and there will exist a small deviation from this straight line trajectory.  Integrating Eq. (\ref{eq:jn}) we find
\begin{equation}
\Delta {1\over c^2}{d\boldsymbol{r}\over dt}=-{1\over E_m}\int_{-\infty}^\infty dt\boldsymbol{\nabla}V_0(r)=-{1\over E_m}\int_{-\infty}^\infty dtV_0'(r)\hat{\boldsymbol{r}}\,,
\end{equation}
so that
\begin{equation}
{1\over c}\theta \simeq {1\over E_m}\int_{-\infty}^\infty dtV_0'(\sqrt{b^2+c^2t^2}){b\over \sqrt{b^2+c^2t^2}}\,,
\end{equation}
where $\theta$ is the bending angle.  Now change variables to $t=bu/c$, yielding
\begin{equation}
{1\over c}\theta\simeq {b\over E_m}\int_{-\infty}^\infty duV_0'(b\sqrt{1+u^2}){1\over \sqrt{1+u^2}}\,.
\end{equation}
That is,
\begin{equation}
\theta_0={b\over E_m}\int_{\infty}^\infty {du\over \sqrt{1+u^2}}V_0'(b\sqrt{1+u^2})={b\over E_m}\int_{-\infty}^\infty {2GE_mMdu\over b^2(1+u^2)^{3\over 2}}={4GM\over b}\,,\label{eq:fg}
\end{equation}
as expected.

The leading correction to Eq. \ref{eq:fg} arises from the one-loop correction to the potential discussed above
\begin{equation}
V_1(r)=-\int{d^3q\over (2\pi)^3}e^{-i\boldsymbol{q}\cdot\boldsymbol{r}}{15\pi^2 G^2M^2E_m\over 2\sqrt{\boldsymbol{q}^2}}=-{15G^2M^2E_m\over 4r^2}\,,
\end{equation}
so the additional bending is given by
\begin{equation}
\theta_1={b\over E_m}\int_{-\infty}^\infty {du\over \sqrt{1+u^2}}V_1'(b\sqrt{1+u^2})={b\over E_m}\int_{-\infty}^\infty {15G^2M^2E_mdu\over 2b^3(1+u^2)^2}={15\pi G^2M^2\over 4b^2}\,,
\end{equation}
in agreement with the result found by the conventional GR method.  We see then that geometrical optics, in which light is treated in a wavelike fashion, provides and interesting alternative way to look at the bending process.

\section{Small-Angle Scattering (Eikonal) Method}

A third approach, ,
discussed in~\cite{Bjerrum-Bohr:2016hpa}, is to look at the bending in terms of a particle interpretation but using quantum mechanics rather than classical physics.\footnote{Note that some previous evaluations which used integration over the calculated cross section accidentally gave the correct answer in the case of the leading contribution but were incorrect at higher order~\cite{Sca72},~\cite{Gol90}.}  In this method we consider a trajectory in terms of a series of small angle high energy scatterings of the photons by the sun. In such a situation the dominant four-momentum transfer is in the transverse spatial directions. For photons travelling in
the $z$-direction we have $p_3=p_1 +q$, so that, squaring, we obtain $0=2E(q_0 -q_z) +q^2$. A similar calculation for the heavy mass
gives $0=-2Mq_0 +q^2$, which tells us that both $q_\pm = q_0 \pm q_z$ are suppressed compared to the transverse components $q^2\sim -\boldsymbol{q}^2_\perp$
by at least a factor of $2E$. This condition on the overall momentum transfer gets reflected in the same stricture for the exchanged gravitons, so that the dominant momentum transfer inside loops is also transverse. (In the effective theory of high energy scattering---Soft Collinear Effective Theory (SCET)---these are called Glauber modes and carry momentum scaling $(k_+,k_-,k_\bot)\sim \sqrt{s}(\lambda^2,\lambda^2,\lambda)$ where $\lambda \sim \sqrt{-t/s}$~\cite{Rot16}.)

The one-graviton amplitude amplitude in this limit was found above to be
\begin{equation}
{\cal M}_0(\boldsymbol{q}_\perp) = -\kappa^2M^2E^2  \frac{1}{\boldsymbol{q}_\perp^2}\,.
\end{equation}
After some work described in Appendix~\ref{sec:gravscat}, the multiple graviton exchanges of this amplitude can be summed, yielding
\begin{equation}
{\cal M}^{(1)}_{tot}(\boldsymbol{q}_\perp) = (4\pi)^2 ME_m   \sum_n \frac{1}{n!} \left({i\over 4}\kappa^2ME_m\right)^n\prod_{i=1}^n \int \frac{d^2\ell_i}{(2\pi)^2}\frac{1}{\boldsymbol{\ell}_i^2} \delta^2(\sum_{j=1}^n \boldsymbol{\ell}_j-\boldsymbol{q}_\perp)\,,\label{eq:kj}
\end{equation}
In order to bring this amplitude into impact parameter space, one defines the two-dimensional Fourier transform, with
impact parameter $\boldsymbol{b}$ being transverse to the initial motion.
\begin{align}
{\cal M}_{tot}(\boldsymbol{b}) =  \int \frac{d^2q_\perp}{(2\pi)^2} e^{i\boldsymbol{q}_\perp\cdot \boldsymbol{b}}~ {\cal M}^{(1)}_{tot}(\boldsymbol{q}_\perp)\,.
\end{align}
We find then
\begin{equation}
{\cal M}_{tot}(\boldsymbol{b})=4ME_m\sum_{n=1}^\infty {1\over n!}(i\chi_0(\boldsymbol{b}))^n=2(s-M^2) \left( e^{i\chi_0(\boldsymbol{b})} -1\right)\,,
\end{equation}
where $\chi_0(\boldsymbol{b})$ is the transverse Fourier transform of the one graviton exchange amplitude
\begin{align}
\chi_0(\boldsymbol{b}) &= \frac{1}{4ME_m}\int \frac{d^2q_\perp}{(2\pi)^2}~e^{i\boldsymbol{q}_\perp\cdot \boldsymbol{b}}~ {\cal M}_0(\boldsymbol{q}_\perp) \ \ \nonumber \\
&= -\frac{\kappa^2ME_m}{4}\int \frac{d^2q_\perp}{(2\pi)^2}~e^{i\boldsymbol{q}_\perp\cdot \boldsymbol{b}} \frac{1}{\boldsymbol{q}_\perp^2}\ \ \nonumber \\
  &=- 4\kappa^2ME_m\left[\frac{1}{d-4} -\log b+...\right] \ \ .
\end{align}
Only the $\log b$ term will be important in calculating the bending angle.  If we now transform back to momentum transfer space via
\begin{equation}
{\cal M}_{tot}(\boldsymbol{q}_\perp)=\int{d^2b\over (2\pi)^2} e^{-i\boldsymbol{q}_\perp\cdot \boldsymbol{b}}{\cal M}_{tot}(\boldsymbol{b})\,,
\end{equation}
and perform the impact parameter space integration via stationary phase methods, we find
\begin{align}
\frac{\partial}{\partial b} \left(|\boldsymbol{q}_\perp| b - \chi_0(b)\right) =0\,.
\end{align}
Since $|\boldsymbol{q}_\perp| =2E_m \sin \frac{\theta}{2}\simeq E_m\theta$ we find
\begin{equation}
\theta_{0} =\frac{1}{E_m}\frac{\partial}{\partial b}\chi_0(b) \,,
\end{equation}
which yields the lowest order result
\begin{equation}
\theta_0={4GM\over b}\,.
\end{equation}
At next to
leading order we require the eikonal phase found from the one-loop correction to massive-massless scattering
\begin{eqnarray}
\chi_1(b)&=&\frac{1}{4ME_m}\int \frac{d^2q_\perp}{(2\pi)^2}~e^{i\boldsymbol{q}_\perp\cdot \boldsymbol{b}}~ {\cal M}_1(\boldsymbol{q}_\perp)\nonumber\\
&=&-\int{d^2q_\perp\over (2\pi)^2}e^{i\boldsymbol{q}_\perp\cdot\boldsymbol{b}}{15\pi^2 G^2M^2E_m\over 2|\boldsymbol{q}_\perp|}=-{15\pi\over 4b}G^2M^2E_m\,.
\end{eqnarray}
The stationary phase calculation then yields the correction
\begin{equation}
\theta_1=\frac{1}{E_m}\frac{\partial}{\partial b}\chi_1(b)={15\pi G^2M^2\over 4b^2}\,,
\end{equation}
which once again agrees with the classical result.

\section{Conclusion}\label{sec:conclusion}

In~\cite{ppa} it was demonstrated that the gravitational interaction of a charged spin-0 or spin-$\frac12$ particle is
greatly simplified by use of factorization, which asserts that the gravitational amplitudes can be written as the product of corresponding electromagnetic amplitudes multiplied by a universal kinematic factor.  In the present work we demonstrated that the same simplification applies when the target particle carries spin-1. Specifically, we evaluated the graviton photo-production and graviton Compton scattering amplitudes explicitly using direct and factorized techniques and demonstrated that they are identical.  However, the factorization methods are {\it enormously} simpler, since they require only electromagnetic calculations and eliminate the need to employ less familiar and more cumbersome tensor quantities. As a result it is now straightforward to include graviton interactions in a quantum mechanics course in order to stimulate student interest and allowing access to various cosmological applications.

We studied the massless limit of the spin-1 system and showed how the use of factorization permits a relatively simple
calculation of graviton-photon scattering, discussing a subtlety in this graviton-photon calculation having to do with the feature that the spin-1 system must change from three to two degrees of freedom when $m\rightarrow 0$ and we explained why the zero mass limit of the spin-1 gravitational Compton scattering amplitude does not correspond to that for photon scattering.  The graviton-photon cross section may possess interesting implications for the attenuation of gravitational waves in the cosmos~\cite{gwc}.  We also calculated the graviton-graviton scattering amplitude.

We discussed the main features of the forward cross-section for each process studied in this paper. Both the Compton and the gravitational Compton scattering have the expected $1/\theta_L^4$ behavior, while graviton photo-production has a different shape that could in principle lead to an interesting new experimental signature of a graviton scattering on matter---$\sim 1/\theta_L^2$. Again this result has potentially intriguing implications for the photo-production of gravitons from
stars~\cite{dqe,pfm}.

Finally, we have reviewed the  evaluation of the classical general relativity
contribution to  the light bending problem---the deviation angle occurring during the passage of a photon by the rim of the sun---in three different ways.  The result in each case was found to be identical
\begin{equation}
\theta={4GM\over b}+{15\pi G^2M^2\over 4b^2}+\ldots
\end{equation}
What is interesting about these results is that they were obtained using apparently very {\it disparate} pictures.

\begin{itemize}
\item [i)] In the first, light is considered from the point of view of photons traversing a classical trajectory in the vicinity of a massive object.
\item [ii)] In the second, the propagation of light is determined by standard geometrical optics, in the presence of an effective index of refraction determined by the effective potential describing the gravitational interaction of massive and massless systems.
\item [iii)] Finally, in the third, standard quantum mechanical scattering methods were used, relating the massive-massless scalar gravitational interaction to the eikonal phase associated with a series of small-angle scatterings.
\end{itemize}

In the latter two cases, it may seem surprising that a method based on a three dimensional Fourier transform, yielding an effective index of refraction, yields results identical to those obtained using small angle scattering theory, involving a two-dimensional eikonal Fourier transform.  However, the equivalence is shown in Appendix~\ref{sec:eik} to result from a simple mathematical identity.\footnote{We note also that by including additional loop contributions, one can also use these methods to evaluate quantum mechanical corrections to the bending angle~\cite{Bjerrum-Bohr:2014zsa, Bjerrum-Bohr:2016hpa}. The origin of such quantum effects can be considered to be zitterbewegung and the feature that the position of the massive scatterer can only be localized to a distance of order its Compton wavelength---$\delta r\sim \hbar/m$.}
What we hope results from this comparison of the various methods is a deeper understanding and illumination of an important general relativistic phenomenon---that of light bending in the presence of a gravitational field.

We close by noting that the same method permits the determination of a quantum effect in the bending
angle, with the result~\cite{Bjerrum-Bohr:2014zsa} (and~\cite{Bai})
\begin{equation}
  \theta_S=  \left(8 bu^S+9-48 \log {b\over 2b_0}\right)\,{\hbar G_N^2M\over \pi b^3}\,,
\end{equation}
where $bu^S$ is a coefficient that depend on the spin $S$ of the
massless particle scattered against the mass stellar object $M$.
The spin dependence of the quantum raises  questions about the
interpretation of the equivalence principle at the quantum mechanical
level~\cite{Bjerrum-Bohr:2015vda,Bjerrum-Bohr:2014zsa,Bjerrum-Bohr:2016hpa} and
strongly suggests additional investigations concerning the nature of the Equivalence
principle at the quantum level that we leave for future work.

\section*{Acknowledgement}

We would like to thank Thibault Damour, Poul H. Damgaard and Gabriele
Veneziano for comments and discussions. The research of P.V. was supported by the ANR grant   reference QFT
ANR 12 BS05 003   01.

\appendix

\section{Equivalence between eikonal and geometrical optics}\label{sec:eik}

\noindent In order to provide a general proof of the equivalence between eikonal and geometrical optics methods, we note that the eikonal phase is, in general, given by
\begin{equation}
\chi(b)=\int{d^2q_\perp\over (2\pi)^2}e^{-i\boldsymbol{q}_\perp\cdot\boldsymbol{b}}{\rm Amp} (\boldsymbol{q}_\perp^2)={1\over 2\pi}\int_0^\infty ds s J_0(b|\boldsymbol{s}|){\rm Amp}(\boldsymbol{s}^2)\,,
\end{equation}
where ${\rm Amp}(s^2)$ is the photon-mass scattering amplitude.  The corresponding contribution to the lightbending angle is then
\begin{equation}
\theta={1\over E_m}{d\over db}\chi(b)=-{1\over 2\pi E_m}\int_0^\infty dss^2J_1(b|\boldsymbol{s}|){\rm Amp}(\boldsymbol{s}^2)\,,
\end{equation}
where we have used $J_0'(x)=-J_1(x)$.\\

On the other hand, in the geometrical optics method, we use the three-dimensional Fourier transform
\begin{equation}
V(r)=\int {d^3q\over (2\pi)^3}e^{-i\boldsymbol{q}\cdot\boldsymbol{r}}{\rm Amp}(\boldsymbol{q}^2)={1\over 2\pi^2}\int_0^\infty dq q^2j_0(|\boldsymbol{q}|r){\rm Amp}(\boldsymbol{q}^2)\,,
\end{equation}
in terms of which the lightbending shift is
\begin{equation}
\theta=-{b\over 2\pi^2E_m}\int_{-\infty}^\infty {du\over \sqrt{1+u^2}}\int_0^\infty dq q^3j_1(|\boldsymbol{q}|b\sqrt{1+u^2}){\rm Amp(\boldsymbol{q}^2)}\,,
\end{equation}
where we have noted $j_0'(x)=-j_1(x)$.  Using $j_1(x)=\sqrt{\pi\over 2x}J_{3\over 2}(x)$
we have
\begin{equation}
\theta=-{b\over 2\pi^2E_m}\int_0^\infty dq q^3G(|\boldsymbol{q}|b){\rm Amp(\boldsymbol{q}^2)}\,,
\end{equation}
where
\begin{equation}
G(|\boldsymbol{q}|b)=\sqrt{\pi\over 2|\boldsymbol{q}|b}\int_{-\infty}^\infty {du\over (1+u^2)^{3\over 4}}J_{3\over 2}(|\boldsymbol{q}|b\sqrt{1+u^2})\,.
\end{equation}
Changing variables to $s=\sqrt{1+u^2}$ so $u=\sqrt{s^2-1}$, then $du=sds/\sqrt{s^2-1}$ and
\begin{equation}
G(|\boldsymbol{q}|b)=\sqrt{\pi\over 2|\boldsymbol{q}|b}\int_{-\infty}^\infty {ds\over \sqrt{s(s^2-1)}}J_{3\over 2}(|\boldsymbol{q}|bs)\,.
\end{equation}
We have~\cite{Gra65}
\begin{equation}
\int_0^\infty {dx J_{3\over 2}(sx)\over \sqrt{x(x^2-1)}}=\sqrt{\pi\over 2s}(J_1(s)-iL_1(-is))\,,
\end{equation}
where $L_1(x)$ is a modified Struve function and satisfies $L_1(x)=L_1(-x)$ while $J_1(x)=-J_1(-x)$ is the usual Bessel function. .  If then we change the integration to the range $-\infty$ to $\infty$, the Struve function disappears and we have
\begin{equation}
G(|\boldsymbol{q}|b)=\sqrt{\pi\over 2|\boldsymbol{q}|b}\cdot\sqrt{2\pi\over |\boldsymbol{q}|b}J_1(|\boldsymbol{q}|b)={\pi\over |\boldsymbol{q}|b}J_1(|\boldsymbol{q}|b)\,,
\end{equation}
whereby
\begin{equation}
\theta=-{1\over 2\pi E_m}\int_0^\infty dqq^2J_1(|\boldsymbol{q}|b){\rm Amp}(\boldsymbol{q}^2),
\end{equation}
which is identical to the eikonal result.

\section{Graviton Scattering Amplitudes}\label{sec:gravscat}

In this appendix we list the independent contributions to the
various graviton scattering amplitudes which must be added in
order to produce the complete and gauge invariant amplitudes quoted in the text.  We
leave it to the (perspicacious) reader to perform the appropriate
additions and to verify the equivalence of the factorized forms shown earlier.\\

\subsection{Graviton Photo-production: spin-1}

\noindent For the case of graviton photoproduction, we find the four contributions, {\it cf.} Fig. 4,
\begin{eqnarray}\!\!\!\!\!\!\!\!\!\!\!\!\!\!\!\!
{\rm Fig. 4(a)}:&& \!\!\!\!\!\!\!\!\!{{\rm Amp}_a(S=1)}={\kappa e\over p_i\cdot k_i}\, \bigg(\epsilon_i\cdot p_i\Big[\epsilon_B^*\cdot\epsilon_A\,\epsilon_f^*\cdot p_f\,\epsilon_f^*\cdot p_f-\epsilon_B^*\cdot k_f\,\epsilon_f^*\cdot p_f\,\epsilon_f^*\cdot\epsilon_A  \nonumber\\
&-&  \epsilon_A\cdot p_f\,\epsilon_f^*\cdot p_f\,\epsilon_f^*\cdot\epsilon_B^*+p_f\cdot k_f\,\epsilon_f^*\cdot\epsilon_A\,\epsilon_f^*\cdot\epsilon_B^*\Big] \nonumber\\
&+& \epsilon_A\cdot\epsilon_i\Big[\epsilon_B^*\cdot k_i\,\epsilon_f^*\cdot p_f\,\epsilon_f^*\cdot p_f-\epsilon_B^*\cdot k_f\,\epsilon_f^*\cdot p_f\,\epsilon_f^*\cdot k_i-p_f\cdot k_i\,\epsilon_f^*\cdot p_f\,\epsilon_f^*\cdot\epsilon_B^*  \nonumber\\
&+&  p_f\cdot k_f\,\epsilon_f^*\cdot k_i\,\epsilon_f^*\cdot\epsilon_B^*\Big] \nonumber\\
&-& \epsilon_A\cdot k_i\,\Big[\epsilon_B^*\cdot\epsilon_i\,\epsilon_f^*\cdot p_f\,\epsilon_f^*\cdot p_f-\epsilon_B^*\cdot k_f\,\epsilon_f^*\cdot p_f\,\epsilon_f^*\cdot\epsilon_i-\epsilon_i\cdot p_f\,\epsilon_f^*\cdot p_f\,\epsilon_f^*\cdot\epsilon_B^*  \nonumber\\
&+&  p_f\cdot k_f\,\epsilon_f^*\cdot\epsilon_i\,\epsilon_f^*\cdot\epsilon_B^*\Big] \nonumber\\
&-& \epsilon_B^*\cdot\epsilon_f^*\,\epsilon_A\cdot\epsilon_i\,\epsilon_f^*\cdot
  p_fp_i\cdot k_i \bigg)\,.
\end{eqnarray}
\begin{eqnarray}\!\!\!\!\!\!\!\!\!\!\!\!\!\!\!\!
{\rm Fig. 4(b)}:&& \!\!\!\!\!\!\!\!\!{{\rm Amp}_b(S=1)}=-{\kappa e\over p_i\cdot k_f}\, \bigg(\epsilon_i\cdot p_f\Big[\epsilon_A\cdot\epsilon_B^*\,\epsilon_f^*\cdot p_i\,\epsilon_f^*\cdot p_i-\epsilon_B^*\cdot p_i\,\epsilon_f^*\cdot p_i\,\epsilon_f^*\cdot\epsilon_A  \nonumber\\
&+&  \epsilon_A\cdot k_f\,\epsilon_f^*\cdot p_i\,\epsilon_f^*\cdot\epsilon_B^*-p_i\cdot k_f\,\epsilon_f^*\cdot\epsilon_A\,\epsilon_f^*\cdot\epsilon_B^*\Big] \nonumber\\
&+& \epsilon_B^*\cdot k_i\Big[\epsilon_A\cdot\epsilon_i\,\epsilon_f^*\cdot p_i\,\epsilon_f^*\cdot p_i-\epsilon_i\cdot p_i\,\epsilon_f^*\cdot p_i\,\epsilon_f^*\cdot\epsilon_A+\epsilon_A\cdot k_f\,\epsilon_f^*\cdot p_i\,\epsilon_f^*\cdot\epsilon_i  \nonumber\\
&-&  p_i\cdot k_f\,\epsilon_f^*\cdot\epsilon_A\,\epsilon_f^*\cdot\epsilon_i\Big] \nonumber\\
&+& \epsilon_i\cdot\epsilon_B^*\Big[\epsilon_A\cdot k_i\,\epsilon_f^*\cdot p_i\,\epsilon_f^*\cdot p_i-p_i\cdot k_i\,\epsilon_f^*\cdot p_i\,\epsilon_f^*\cdot\epsilon_A+\epsilon_A\cdot k_f\,\epsilon_f^*\cdot p_i\,\epsilon_f^*\cdot k_i  \nonumber\\
&-&  p_i\cdot k_f\,\epsilon_f^*\cdot\epsilon_A\,\epsilon_f^*\cdot k_i\Big] \nonumber\\
&-& \epsilon_A\cdot\epsilon_f^*\,\epsilon_f^*\cdot p_i\,\epsilon_B^*\cdot\epsilon_i\,p_i\cdot k_f \bigg)\,.
\end{eqnarray}
\begin{eqnarray}
{\rm Fig. 4(c)}:&&\!\!\!\!\!\!\!\!\!{\rm Amp}_c(S=1)=\kappa \,e \bigg(\epsilon_f^*\cdot\epsilon_i(\epsilon_B^*\cdot\epsilon_A\,\epsilon_f^*\cdot (p_f+p_i)-\epsilon_A\cdot p_f\,\epsilon_B^*\cdot\epsilon_f^*
-\epsilon_B^*\cdot p_i\,\epsilon_A\cdot\epsilon_f^*) \nonumber\\
&-& \epsilon_B^*\cdot\epsilon_f^*\,\epsilon_A\cdot\epsilon_i\,\epsilon_f^*\cdot p_i-\epsilon_A\cdot\epsilon_f^*\,\epsilon_B^*\cdot\epsilon_i\,\epsilon_f^*\cdot p_f+\epsilon_f^*\cdot\epsilon_A\,\epsilon_f^*\cdot\epsilon_B^*\,\epsilon_i\cdot(p_f+p_i) \bigg)\,,
\end{eqnarray}
and finally, the photon pole contribution
\begin{eqnarray}\!\!\!\!\!\!\!\!\!\!\!\!\!\!\!\!\!\!\!\!\!\!\!\!\!\!\!\!\!\!\!\!\!\!\!\!\!\!\!\!\!\!\!\!\!\!\!\!\!\!\!\!\!\!\!\!\!\!\!\!\!\!\!\!\!\!\!\!\!\!\!\!\!\!\!\!\!\!\!\!\!\!\!\!\!\!\!\!\!
{\rm Fig. 4(d)}:&&\!\!\!\! {{\rm Amp_d}(S=1)}=-{e\,\kappa\over 2k_f\cdot k_i}\nonumber\\
&\times& \bigg[\epsilon_B^*\cdot\epsilon_A\Big[\epsilon_f^*\cdot(p_f+p_i)(k_f\cdot
k_i\epsilon_f^*\cdot\epsilon_i-\epsilon_f^*\cdot k_i\,\epsilon_i\cdot k_f)  \nonumber\\
&+&  \epsilon_f^*\cdot
k_i(\epsilon_f^*\cdot\epsilon_i\, k_i\cdot(p_i+p_f)-\epsilon_f^*\cdot
k_i\epsilon_i\cdot(p_f+p_i))\Big] \nonumber\\
&-& 2\epsilon_B^*\cdot p_i\Big[\epsilon_f^*\cdot\epsilon_A\,(k_f\cdot
k_i\,\epsilon_f^*\cdot\epsilon_i-\epsilon_f^*\cdot
k_i\,\epsilon_i\cdot k_f)  \nonumber\\
&+&  \epsilon_f^*\cdot
k_i\,(\epsilon_f^*\cdot\epsilon_i\, \epsilon_A\cdot k_i-\epsilon_f^*\cdot
k_i\,\epsilon_i\cdot\epsilon_A)\Big] \nonumber\\
&-& 2\epsilon_A\cdot p_f\Big[\epsilon_f^*\cdot\epsilon_B^*(k_f\cdot
k_i\,\epsilon_f^*\cdot\epsilon_i-\epsilon_f^*\cdot
k_i\,\epsilon_i\cdot k_f)  \nonumber\\
&+&  \epsilon_f^*\cdot
k_i\,(\epsilon_f^*\cdot\epsilon_i \,\epsilon_B^*\cdot k_i-\epsilon_f^*\cdot
k_i\,\epsilon_i\cdot\epsilon_B^*)\Big] \bigg]\,.
\end{eqnarray}

\subsection{Gravitational Compton Scattering: spin-1}

\noindent In the case of gravitational Compton scattering---Figure~\ref{fig:gravcomp}---we have the four contributions
\begin{align}
{\rm Fig. 5(a)}:& {\rm Amp}_a(S=1)=\kappa^2\,{1\over 2p_i\cdot
k_i}\bigg[(\epsilon_i\cdot p_i)^2 (\epsilon_f^*\cdot
p_f)^2\epsilon_A\cdot\epsilon_B^*\nonumber\\
&-(\epsilon_f^*\cdot p_f)^2\epsilon_i\cdot p_i (\epsilon_A\cdot
k_i\epsilon_B^*\cdot \epsilon_i+
\epsilon_A\cdot\epsilon_i\,\epsilon_B^*\cdot p_i)\nonumber\\
&-(\epsilon_i\cdot p_i)^2\epsilon_f^*\cdot p_f\,
(\epsilon_B^*\cdot\epsilon_f^*\,\epsilon_A\cdot
p_f+\epsilon_B^*\cdot
k_f\,\epsilon_A\cdot\epsilon_f^*)\nonumber\\
&+\epsilon_i\cdot p_i\,\epsilon_f^*\cdot p_f\,\epsilon_i\cdot
p_f\,\epsilon_A\cdot
k_i\,\epsilon_B^*\cdot\epsilon_f^*+\epsilon_i\cdot
p_i\,\epsilon_f^*\cdot p_f\,\epsilon_f^*\cdot p_i\,\epsilon_A\cdot
\epsilon_i\,\epsilon_B^*\cdot k_f\nonumber\\
&+(\epsilon_f^*\cdot
p_f)^2\epsilon_B^*\cdot\epsilon_i\,\epsilon_A\cdot\epsilon_i\,
p_i\cdot k_i+(\epsilon_i\cdot
p_i)^2\epsilon_B^*\cdot\epsilon_f^*\,\epsilon_A\cdot\epsilon_f^*\,
p_f\cdot k_f\nonumber\\
&+\epsilon_i\cdot p_i\,\epsilon_f^*\cdot p_f\,(\epsilon_A\cdot
k_i\,\epsilon_B^*\cdot k_f\,\epsilon_i\cdot
\epsilon_f^*+\epsilon_B^*\cdot\epsilon_f^*\,\epsilon_A\cdot\epsilon_i\,p_i\cdot
p_f)\nonumber\\
&-\epsilon_i\cdot p_i\,\epsilon_f^*\cdot p_i\,\epsilon_B^*\cdot
\epsilon_f^*\,\epsilon_A\cdot\epsilon_i\,p_f\cdot
k_f-\epsilon_f^*\cdot p_f\,\epsilon_i\cdot
p_f\,\epsilon_A\cdot\epsilon_i\,\epsilon_B^*\cdot\epsilon_f^*\,p_i\cdot
k_i\nonumber\\
&-\epsilon_i\cdot p_i\,\epsilon_A\cdot k_i\,\epsilon_B^*\cdot
\epsilon_f^*\,\epsilon_f^*\cdot\epsilon_i\,p_f\cdot
k_f-\epsilon_f^*\cdot p_f\,\epsilon_B^*\cdot
k_f\,\epsilon_A\cdot\epsilon_i\,\epsilon_i\cdot\epsilon_f^*\,p_i\cdot
k_i\nonumber\\
&+\epsilon_A\cdot\epsilon_i\,\epsilon_B^*\cdot\epsilon_f^*\,p_i\cdot
k_i\,p_f\cdot
k_f\,\epsilon_i\cdot\epsilon_f^*-m^2\epsilon_B^*\cdot\epsilon_f^*\,
\epsilon_A\cdot\epsilon_i\,\epsilon_f^*\cdot
p_f\,\epsilon_i\cdot p_i\bigg]\,.
\end{align}
\begin{align}\!\!\!\!\!\!\!\!\!\!\!\!\!\!
{\rm Fig. 5(b)}:& {\rm Amp}_b(S=1)=-\kappa^2{1\over 2p_i\cdot
k_f}\bigg[\big(\epsilon_f^*\cdot p_i\big)^2 \big(\epsilon_i\cdot
p_f\big)^2\epsilon_A\cdot\epsilon_B^*\nonumber\\
&+\big(\epsilon_i\cdot p_f\big)^2\epsilon_f^*\cdot p_i
\big(\epsilon_A\cdot k_f\,\epsilon_B^*\cdot\epsilon_f^*
-\epsilon_A\cdot\epsilon_f^*\,\epsilon_B^*\cdot p_i\big)\nonumber\\
&+\big(\epsilon_f^*\cdot p_i\big)^2\epsilon_i\cdot p_f \big(\epsilon_B^*\cdot
k_i\epsilon_A\cdot\epsilon_i-\epsilon_B^*\cdot\epsilon_i\,\epsilon_A\cdot p_f\big)\nonumber\\
&-\epsilon_f^*\cdot p_i\,\epsilon_i\cdot p_f\,\epsilon_f^*\cdot
p_f\,\epsilon_A\cdot
k_f\,\epsilon_B^*\cdot\epsilon_i-\epsilon_f^*\cdot
p_i\,\epsilon_i\cdot p_f\,\epsilon_i\cdot p_i\,\epsilon_A\cdot
\epsilon_f^*\,\epsilon_B^*\cdot k_i\nonumber\\
&-\big(\epsilon_i\cdot
p_f\big)^2\epsilon_B^*\cdot\epsilon_f^*\,\epsilon_A\cdot\epsilon_f^*\,
p_i\cdot k_f-\big(\epsilon_f^*\cdot
p_i\big)^2\epsilon_B^*\cdot\epsilon_i\,\epsilon_A\cdot\epsilon_i\,
p_f\cdot k_i\nonumber\\
&+\epsilon_f^*\cdot p_i\,\epsilon_i\cdot p_f\big(\epsilon_A\cdot
k_f\,\epsilon_B^*\cdot k_i\,\epsilon_i\cdot
\epsilon_f^*+\epsilon_B^*\cdot\epsilon_i\,\epsilon_A\cdot\epsilon_f^*\,p_i\cdot
p_f\big)\nonumber\\
&+\epsilon_f^*\cdot p_i\,\epsilon_i\cdot p_i\,\epsilon_B^*\cdot
\epsilon_i\,\epsilon_A\cdot\epsilon_f^*\,p_f\cdot k_i+\epsilon_i\cdot
p_f\,\epsilon_f^*\cdot
p_f\,\epsilon_A\cdot\epsilon_f^*\,\epsilon_B^*\cdot\epsilon_i\,p_i\cdot
k_f\nonumber\\
&-\epsilon_f^*\cdot p_i\,\epsilon_A\cdot k_f\,\epsilon_B^*\cdot
\epsilon_i\,\epsilon_i\cdot\epsilon_f^*\,p_f\cdot k_i-\epsilon_i\cdot
p_f\,\epsilon_B^*\cdot
k_i\,\epsilon_A\cdot\epsilon_f^*\,\epsilon_f^*\cdot\epsilon_i\,p_i\cdot
k_f\nonumber\\
&+\epsilon_A\cdot\epsilon_f^*\,\epsilon_B^*\cdot\epsilon_i\,p_i\cdot
k_f\,p_f\cdot
k_i\,\epsilon_i\cdot\epsilon_f^*-m^2\epsilon_B^*\cdot\epsilon_i\,
\epsilon_A\cdot\epsilon_f^*\,\epsilon_i\cdot
p_f\,\epsilon_f^*\cdot p_i\bigg]\,.
\end{align}
\begin{align}
{\rm Fig. 5(c)}:&{\rm Amp}_c(S=1)=-{\kappa^2\over 4}\,
\bigg[\big(\epsilon_i\cdot\epsilon_f^*\big)^2\big(m^2-p_i\cdot
p_f\big)\epsilon_A\cdot\epsilon_B^*\nonumber\\
&+\epsilon_A\cdot
p_f\,\epsilon_B^*\cdot
p_i\,\big(\epsilon_i\cdot\epsilon_f^*\big)^2
+\epsilon_i\cdot p_i\,\epsilon_f^*\cdot
p_f\,\big(2\epsilon_i\cdot\epsilon_f^*\,\epsilon_A\cdot\epsilon_B^*-
2\epsilon_A\cdot\epsilon_2\,\epsilon_B^*\cdot\epsilon_1\big)\nonumber\\
&+\epsilon_i\cdot p_f\,\epsilon_f^*\cdot
p_i\,\big(2\epsilon_i\cdot\epsilon_f^*\,\epsilon_A\cdot\epsilon_B^*-
2\epsilon_A\cdot\epsilon_i\,\epsilon_B^*\cdot\epsilon_f^*\big)\nonumber\\
&+2\epsilon_i\cdot p_i\,\epsilon_1\cdot
p_f\,\epsilon_A\cdot\epsilon_f^*\,\epsilon_B^*\cdot\epsilon_f^*+2\epsilon_f^*\cdot
p_f\,\epsilon_f^*\cdot
p_i\,\epsilon_A\cdot\epsilon_i\,\epsilon_B^*\cdot\epsilon_i\nonumber\\
&-2\epsilon_i\cdot p_i\,\epsilon_i\cdot\epsilon_f^*\,\epsilon_A\cdot
p_f\,\epsilon_B^*\cdot\epsilon_f^*-2\epsilon_f^*\cdot
p_f\,\epsilon_i\cdot\epsilon_f^*\,\epsilon_A\cdot\epsilon_i\,\epsilon_f^*\cdot
p_i\nonumber\\
&-2\epsilon_i\cdot
p_f\,\epsilon_i\cdot\epsilon_f^*\,\epsilon_A\cdot\epsilon_f^*\,\epsilon_B^*\cdot
p_i-2\epsilon_f^*\cdot
p_i\,\epsilon_i\cdot\epsilon_f^*\,\epsilon_B^*\cdot\epsilon_i\,\epsilon_A\cdot
p_f\nonumber\\
&-2\big(m^2-p_f\cdot
p_i\big)\epsilon_i\cdot\epsilon_f^*\big(\epsilon_A\cdot\epsilon_i\,\epsilon_B^*\cdot\epsilon_f^*
+\epsilon_A\cdot\epsilon_f^*\,\epsilon_B^*\cdot\epsilon_i\big)\bigg]\,,
\end{align}

and finally the (lengthy) graviton pole contribution is
\begin{align}\!\!\!\!\!\!\!\!\!\!\!\!\!\!\!\!\!
 {\rm 5(d)}:&{\rm Amp}_d(S=1)=-{\kappa^2\over 16\,k_i\cdot
k_f}\bigg[\epsilon_B^*\cdot\epsilon_A\Big[\big(\epsilon_i\cdot\epsilon_f^*\big)^2\Big[4k_i\cdot
p_i\,p_f\cdot
k_i+4k_f\cdot p_i\,k_f\cdot p_f\nonumber\\ \nonumber
&-2\big(p_i\cdot k_i\,p_f\cdot k_f+p_f\cdot k_i\,p_i\cdot k_f\big)+6p_i\cdot
p_f\,k_i\cdot k_f\Big]+4\Big[\big(\epsilon_i\cdot k_f\big)^2\epsilon_f^*\cdot p_f\,\epsilon_f^*\cdot
p_i\\ \nonumber 
&+\big(\epsilon_f^*\cdot k_i\big)^2\epsilon_i\cdot p_i\,\epsilon_i\cdot
p_f
+\epsilon_i\cdot k_f\,\epsilon_f^*\cdot k_i\big(\epsilon_i\cdot
p_i\,\epsilon_f^*\cdot p_f+\epsilon_i\cdot p_f\,\epsilon_f^*\cdot
p_i\big)\Big]\nonumber\\
&-4\epsilon_i\cdot\epsilon_f^*\Big[\epsilon_i\cdot
k_f\big(\epsilon_f^*\cdot
p_i\, p_f\cdot k_f+\epsilon_f^*\cdot p_f\,k_f\cdot p_i\big)
+\epsilon_f^*\cdot k_i\big(\epsilon_i\cdot p_i\,p_f\cdot
k_i+\epsilon_i\cdot p_f\,p_i\cdot k_i\big)\Big]\nonumber\\
&-4k_i\cdot k_f\,\epsilon_i\cdot\epsilon_f^*\big(\epsilon_i\cdot
p_i\,\epsilon_f^*\cdot p_f+\epsilon_i\cdot p_f\,\epsilon_f^*\cdot
p_i\big)-4p_i\cdot p_f\,\epsilon_i\cdot\epsilon_f^*\,\epsilon_i\cdot
k_f\,\epsilon_f^*\cdot k_i\Big]\nonumber\\
&-\big(p_i\cdot p_f\,\epsilon_B^*\cdot\epsilon_A-\epsilon_B^*\cdot
p_i\,\epsilon_A\cdot p_f\big)\Big[10\big(\epsilon_i\cdot\epsilon_f^*\big)^2k_i\cdot
k_f+4\epsilon_i\cdot\epsilon_f^*\,\epsilon_i\cdot
k_f\,\epsilon_f^*\cdot
k_i\nonumber\\
&-4\big(\epsilon_i\cdot\epsilon_f^*\big)^2k_i\cdot
k_f-8\epsilon_i\cdot\epsilon_f^*\,\epsilon_i\cdot
k_f\,\epsilon_f^*\cdot
k_i\Big]+\big(p_i\cdot
p_f-m^2\big)\Big[\big(\epsilon_i\cdot\epsilon_f^*\big)^2\big(4\epsilon_A\cdot
k_i\,\epsilon_B^*\cdot k_i\nonumber \\ \nonumber&+4\epsilon_A\cdot k_f\,\epsilon_B^*\cdot
k_f-2\big(\epsilon_A\cdot k_i\,\epsilon_B^*\cdot k_f+\epsilon_A\cdot
k_f\,\epsilon_B^*\cdot k_i\big)+6\epsilon_B^*\cdot\epsilon_A\,k_i\cdot
k_f\big)\nonumber\\
&+4\Big[\big(\epsilon_i\cdot
k_f\big)^2\epsilon_A\cdot\epsilon_f^*\,\epsilon_B^*\cdot\epsilon_f^*+\big(\epsilon_f^*\cdot
k_i\big)^2\epsilon_A\cdot\epsilon_i\,\epsilon_B^*\cdot\epsilon_i
+\epsilon_i\cdot k_f\,\epsilon_f^*\cdot
k_f\big(\epsilon_A\cdot\epsilon_i\,\epsilon_B^*\cdot\epsilon_f^*\nonumber \\ \nonumber
&+\epsilon_A\cdot\epsilon_f^*\,\epsilon_B^*\cdot\epsilon_i\big)\Big]-4\epsilon_i\cdot\epsilon_f^*\Big[\epsilon_i\cdot
k_f\big(\epsilon_A\cdot\epsilon_f^*\,\epsilon_B^*\cdot
k_f+\epsilon_B^*\cdot\epsilon_f^*\,\epsilon_A\cdot k_f\big)\nonumber\\
&+\epsilon_f^*\cdot
k_i\big(\epsilon_A\cdot\epsilon_i\,\epsilon_B^*\cdot
k_i+\epsilon_B^*\cdot\epsilon_i\,\epsilon_A\cdot k_i\big)+k_i\cdot
k_f\big(\epsilon_A\cdot\epsilon_i\,\epsilon_B^*\cdot\epsilon_f^*+
\epsilon_B^*\cdot\epsilon_i\,\epsilon_A\cdot\epsilon_f^*\big)\nonumber\\ \nonumber&+\epsilon_A\cdot\epsilon_B^*\,\epsilon_i\cdot
k_f\,\epsilon_f^*\cdot k_i\Big]\Big]
-2\epsilon_A\cdot
p_f\Big[\big(\epsilon_f^*\cdot\epsilon_i\big)^2\Big[2\epsilon_B^*\cdot k_i\,p_i\cdot
k_i+2\epsilon_B^*\cdot k_f\,p_i\cdot k_f\nonumber \\ \nonumber & +3\epsilon_B^*\cdot
p_i\,k_i\cdot k_f-\big(\epsilon_B^*\cdot k_i\,p_i\cdot k_f+\epsilon_B^*\cdot
k_f\,p_i\cdot k_i\big)\Big]+2\big(\epsilon_i\cdot
k_f\big)^2\epsilon_B^*\cdot\epsilon_f^*\,\epsilon_f^*\cdot
p_i\nonumber\\ \nonumber&+2\big(\epsilon_f^*\cdot
k_i\big)^2\epsilon_B^*\cdot\epsilon_i\,\epsilon_i\cdot p_i+2\epsilon_i\cdot k_f\epsilon_f^*\cdot
k_i\big(\epsilon_B^*\cdot\epsilon_i\,\epsilon_f^*\cdot
p_i+\epsilon_i\cdot
p_i\,\epsilon_B^*\cdot\epsilon_f^*\big)\nonumber\\
&-2\epsilon_i\cdot\epsilon_f^*\Big[\epsilon_i\cdot
k_f\big(\epsilon_B^*\cdot\epsilon_f^*\,p_i\cdot k_f+\epsilon_f^*\cdot
p_i\,\epsilon_B^*\cdot k_f\big)+\epsilon_f^*\cdot k_i\big(\epsilon_B^*\cdot\epsilon_i\,p_i\cdot
k_i+\epsilon_B^*\cdot k_i\,\epsilon_i\cdot p_i\big)\Big]\nonumber\\
&-2k_i\cdot
k_f\,\epsilon_i\cdot\epsilon_f^*\big(\epsilon_B^*\cdot\epsilon_i\,\epsilon_f^*\cdot
p_i+\epsilon_B^*\cdot\epsilon_f^*\,\epsilon_i\cdot p_i\big)
-2\epsilon_B^*\cdot p_i\,\epsilon_i\cdot\epsilon_f^*\,\epsilon_i
\cdot k_f\,\epsilon_f^*\cdot k_i\Big]\nonumber\\
&-2\epsilon_B^*\cdot
p_i\Big[\big(\epsilon_f^*\cdot\epsilon_i\big)^2\Big[2\epsilon_A\cdot k_i\,p_f\cdot
k_i+2\epsilon_A\cdot k_f\,p_f\cdot k_f+3\epsilon_A\cdot p_f\,k_i\cdot
k_f\nonumber\\
&-\big(\epsilon_A\cdot k_i\,p_f\cdot k_f+\epsilon_A\cdot k_f\,p_f\cdot
k_f)\Big]+2\big(\epsilon_i\cdot
k_f\big)^2\epsilon_A\cdot\epsilon_f^*\,\epsilon_f^*\cdot
p_f+2\big(\epsilon_f^*\cdot
k_i\big)^2\epsilon_A\cdot\epsilon_i\,\epsilon_i\cdot p_f\nonumber\\
&+2\epsilon_i\cdot k_f\,\epsilon_f^*\cdot
k_i\big(\epsilon_A\cdot\epsilon_i\,\epsilon_f^*\cdot p_f+\epsilon_i\cdot
p_f\,\epsilon_A\cdot\epsilon_f^*-2\epsilon_i\cdot\epsilon_f^*\Big[\epsilon_i\cdot
k_f\big(\epsilon_A\cdot\epsilon_f^*\,p_f\cdot k_f\nonumber\\ \nonumber&+\epsilon_f^*\cdot
p_f\,\epsilon_A\cdot k_f\big)
+\epsilon_f^*\cdot k_i\big(\epsilon_A\cdot\epsilon_i\,p_f\cdot
k_i\nonumber +\epsilon_A\cdot k_i\,\epsilon_i\cdot p_f\big)\Big]\nonumber\\
&-2k_i\cdot
k_f\epsilon_i\cdot\epsilon_f^*\big(\epsilon_A\cdot\epsilon_i\,\epsilon_f^*\cdot
p_f+\epsilon_A\cdot\epsilon_f^*\,\epsilon_i\cdot p_f\big)
-2\epsilon_A\cdot p_f\,\epsilon_i\cdot\epsilon_f^*\,\epsilon_i\cdot
k_f\,\epsilon_f^*\cdot k_i\Big]\bigg]\,.
\end{align}

\end{document}